\documentclass[structabstract]{aa} % for printer 
\usepackage{graphicx}
\usepackage{txfonts}
\usepackage{subfigure}
\begin{document}

\title{Multi-frequency properties of synthetic blazar radio light
  curves within the shock-in-jet scenario}
               
\author{C. M. Fromm\inst{1}, L. Fuhrmann\inst{1} and M. Perucho\inst{2,3}}
\institute{Max-Planck-Institut f\"ur Radioastronomie Auf dem H\"ugel 69, D-53121 Bonn, Germany \\
\email{cfromm@mpifr.de, lfuhrmann@mpifr.de, perucho@uv.es}
\and Departament d'Astronomia i Astrofisica, Universitat de Valencia C/Dr. Moliner 50, 46100 Burjassot, Valencia, Spain
\and Observatori Astron\`omic, Parc Cient\'{\i}fic, Universitat de Val\`encia, C/ Catedr\`atic Jos\'e Beltr\'an 2, E-46980 Paterna (Val\`encia), Spain }

%Current address: Deutsches Zentrum f\"ur Luft- und Raumfahrt, Institut f\"ur Raumfahrtantriebe, Langer Grund, D-74239 Hardthausen, Germany}
%\date{Received September 15, 1996; accepted March 16, 1997}

% \abstract{}{}{}{}{} 
% 5 {} token are mandatory
 
\abstract
{Blazars are among the most powerful extragalactic objects, as a
  sub-class of active galactic nuclei. They launch relativistic jets
  and their emitted radiation shows strong variability across the
  entire electro-magnetic spectrum. The mechanisms producing the
  variability are still controversial and different models have been
  proposed to explain the observed variations in multi-frequency
  blazar light curves.}
{We investigate the capabilities of the classical shock-in-jet model
  to explain and reconstruct the observed evolution of flares in the
  turnover frequency -- turnover flux density ($\nu_\mathrm{m}$--$S_\mathrm{m}$) plane
  and their frequency-dependent light curve parameters.  With a
  detailed parameter space study we provide the framework for future,
  detailed comparisons of observed flare signatures with the
  shock-in-jet scenario.}
{Based on the shock model we compute synthetic single-dish light
  curves at different radio frequencies (2.6 to 345\,GHz) and for
  different physical conditions in a conical jet (e.g. magnetic field
  geometry and Doppler factor). From those we extract the slopes of
  the different energy loss stages within the $\nu_\mathrm{m}$--$S_\mathrm{m}$ plane and
  deduce the frequency-dependence of different light curve parameters
  such as flare amplitude, time scale and cross-band delays.}
{The evolution of the Doppler factor along the jet has the largest
  influence on the evolution of the flare and on the
  frequency-dependent light curve parameters. The synchrotron stage
  can be hidden in the Compton or in the adiabatic stage, depending
  mainly on the evolution of the Doppler factor, which makes it
  difficult to detect its signature in observations. In addition, we
  show that the time lags between different frequencies can be used as
  an efficient tool to better constrain the physical properties of
  these objects.}
{}

\keywords{Galaxies: active, blazars, Galaxies: jets, Radiation mechanisms: non-thermal, Methods: analytical}
\authorrunning{C. M. Fromm et al.}   
\titlerunning{Properties of synthetic blazar radio light curves within the shock-in-jet scenario}

\maketitle
%
%________________________________________________________________

\section{Introduction}
Blazars (Flat-Spectrum Radio Quasars and BL\,Lacertae objects) are a
sub-class of powerful active galactic nuclei (AGN). They exhibit
extreme phenomenological characteristics mostly attributed to small
viewing angles with respect to their jet axis
($\lesssim$\,20$^{\circ}$). Among those, rapid broad band flux density
and polarization variability, fast superluminal motion, high degree of
polarization, and a broad-band, double-humped spectral energy
distribution (SED) are commonly observed
\citep[e.g.][]{1999APh....11..159U}.

The origin of the rapid (time scales of days/weeks to months/years)
variability of the synchrotron and high-energy branch of blazar SEDs
is still a matter of debate. The observed rapid variability probes
spatial scales often inaccessible even to interferometric imaging and
has been explained in terms of e.g., relativistic shock-in-jet models
\citep[e.g.,][]{1985ApJ...298..114M,1992A&A...254...80V,1992A&A...254...71V,1994ApJ...437...91S,2000A&A...361..850T}
or colliding relativistic plasma shells
\citep[e.g.,][]{2001MNRAS.325.1559S,2004A&A...421..877G,2007A&A...466...93M,2014MNRAS.438.1856R}.
Quasi-periodicities on time scales of months to years indicate
systematic changes in the beam orientation
\citep{1992A&A...255...59C}, possibly related to binary black hole
systems, magneto-hydrodynamic instabilities in the accretion disks
and/or helical/precessing jets
\citep[e.g.,][]{1980Natur.287..307B,1999A&A...347...30V}. {\citet{2012ApJ...749...55P} and \citet{2013AJ....146..120L} have suggested that the radio emission from jets could be emitted from a portion of the jet cross-section that is brighter due to instability structures within the jet. \citet{2013AJ....146..120L} also pointed out that radio-components that could be associated to shocks travelling through the jet are generally non-ballistic. This fact is revealed at parsec-scales, and there is no evidence for this behaviour to be extended to more compact scales, where the flares are bright at higher frequencies. If that is the case, the changes in the direction of propagation of the bright features associated to flares would translate in variable flux associated to variations in the Doppler factor.}{
  Different} variability models predict different frequency
dependencies for various observables in the time domain, such as
variability amplitudes, characteristic time scales, or cross-band
delays. According to these predictions, different spectral
characteristics and evolution are expected for quantities such as the
radio spectral index and the turnover frequency, $\nu_\mathrm{m}$, and turnover
flux density, $S_\mathrm{m}$. Previous detailed studies support shocks as the
origin of the observed radio variability in cm/mm band blazar light
curves: individual source and flare studies in the {time and/or
  spectral domain}
\citep[e.g.][]{1985ApJ...298..114M,2000A&A...361..850T,
  2011A&A...531A..95F,2013MNRAS.428.2418O,2013A&A...552A..11R}, as
well as studies of larger samples and/or many individual flares
\citep[e.g.][]{1992A&A...254...71V,1994ApJ...437...91S,
  2008A&A...485...51H,2011arXiv1111.6992A,fuhrmann2014a}
often show an overall good agreement of the observed flare signatures
with the shock-in-jet scenario. However, in order to test the
different variability models and their multi-frequency predictions in
detail, broad band radio AGN/blazar monitoring programs of larger
samples are of uttermost importance in providing the necessary
constraints for understanding the origin of variability and energy
production.  The \textit{Fermi}-GST AGN Multi-frequency Monitoring
Alliance { (F-GAMMA
  program)}\footnote{http://www3.mpifr-bonn.mpg.de/div/vlbi/fgamma/fgamma.html}
\citep{Fuhrmann:2007bh,2010ASPC..427..289A,2014MNRAS.441.1899F}, 
for instance, is providing such detailed data sets and observational
parameters since 2007, monitoring contemporaneously the total flux
density, polarisation and spectral evolution of about 60 {\it Fermi}
{$\gamma$-ray} blazars at three radio observatories with a cadence of
about one month. The overall frequency range spans from 2.64 to
345\,GHz (110\,mm to 0.8\,mm) using the Effelsberg 100-m, IRAM 30-m,
and APEX 12-m telescopes at a total of 11 frequency bands.

In order to provide the framework for future, detailed comparisons of
theoretical models and F-GAMMA radio monitoring data, we focus in the
current work on the shock-in-jet scenario and study its predictions of
frequency-dependent light curve parameters as a function of different
physical conditions. We perform a parameter space study for conical
jets within the shock-in-jet model, where we use different scenarios
for the evolution of the magnetic field with the jet radius
  $R$, $B\,\propto\,R^{-b}$, the spectral slope, $s$, of the relativistic
electron distribution with energy $E$, $N\,\propto\,E^{-s}$, and
the Doppler factor along the jet, $\delta\,\propto\,R^{-d}$.
For each set of jet conditions given by $b$, $s$ and $d$, we compute
the slopes of the different energy loss stages, namely the Compton, 
synchrotron and adiabatic stage and based on these slopes we calculate
synthetic single-dish light curves.
From the synthetic light curves we extract frequency-dependent single-dish light
curve parameters, i.e. flare amplitudes, time scales and cross-band  
delays \citep[][]{1985ApJ...298..114M,2011A&A...531A..95F}. Moreover, we
investigate in an additional model the impacts of second order Compton
scattering on those parameters \citep{2000ApJ...533..787B}.

The paper is structured as follows. In Section~2 we review the
standard shock-in-jet model, the parameter space for the physical
conditions in the jet, and our analysis applied to the synthetic light
curves.  We present the results of this analysis in Section~3,
Section~4 includes the discussion, and in Section~5 we list the main
conclusions and outlook of this paper.

\section{The shock model}\label{tsm}
We used the classical shock-in-jet model to compute the spectral
evolution during a flare within the mm to cm wavelength regime
{\citep{1985ApJ...298..114M,2000A&A...361..850T,2011A&A...531A..95F}}. This model assumes
that a shock wave is propagating through a conical jet (direct
proportionality between the jet radius $R$ and the distance along the
jet $r$, $R\propto r^{\rho}$, with $\rho=1$) and accelerates
relativistic particles at the shock front. These particles travel
behind the shock front and { loose} their energy due to different
energy loss mechanisms, {namely} Compton, synchrotron and
adiabatic losses. The {energy} distribution of the relativistic {electrons}
is assumed to be a power-law {$N(E)=KE^{-s}$}, with $K$ the normalisation
coefficient, $E$ the energy, and $s$ the spectral slope. The evolution
of the physical parameters, e.g., the magnetic field, $B$, {the
normalisation coefficient $K$}, and the Doppler factor, $\delta$, in
the bulk flow are parametrized by power-laws
\begin{equation}
B\propto R^{-b}\quad K\propto R^{-k} \quad \delta\propto R^{-d}.
\label{eq:bkr}
\end{equation}
Given these dependencies and the standard {synchrotron self-absorption}
{(SSA)} theory, the evolution of the turnover-frequency, $\nu_\mathrm{m}$,
and the turnover flux density, $S_\mathrm{m}$, can be written as a power-law:

\begin{equation}
S_\mathrm{m}\propto \nu_\mathrm{m}^{\epsilon_i},
\label{smvm}
\end{equation}
where the exponent $\epsilon_i=f_i/n_i$ is given below, and the
subscript $i=1$ corresponds to the Compton, $i=2$ to the synchrotron
and $i=3$ to the adiabatic stage, with:
\begin{eqnarray}
n_1&=&-(b+1)/4-d(s+3)/(s+5) \label{an1}\\
n_2&=&-[2k+b(s-1)+d(s+3)]/(s+5)\\
n_3&=&-[2(k-1)+(b+d)(s+2)]/(s+4)\label{n3}\\
f_1&=&[(11-b)/8]-[d(3s+10)/(s+5)]\\
f_2&=&2-[5k+b(2s-5)+d(3s+10)]/(s+5)\\
f_3&=&[2s+13-5k-b(2s+3)-d(3s+7)]/(s+4)\label{f3},
\end{eqnarray}
{In the following, we label $\epsilon_1=\epsilon_\mathrm{C(ompton)}$, $\epsilon_2=\epsilon_\mathrm{S(ynchrotron)}$ and $\epsilon_3=\epsilon_\mathrm{A(diabatic)}$} 

In addition to the evolution of the flare in the turnover
frequency--turnover flux density $\left(\nu_\mathrm{m}-S_\mathrm{m}\right)$ plane, the
temporal evolution of the turnover frequency and the turnover flux
density can be computed:
\begin{eqnarray}
\nu_\mathrm{m}\propto t^{n_i/\xi} \label{vmt}\\
S_\mathrm{m}\propto t^{f_i/\xi} \label{smt},
\end{eqnarray}
where $\xi=(2d\rho+1)/\rho$ \citep[see, e.g.,][for a detailed
derivation]{2011A&A...531A..95F}. The proportionality constants in
Eq.~\ref{smvm}, Eq.~\ref{vmt}, and Eq.\ref{smt} can be calculated by
using the turnover frequency and the turnover flux density at the end
of the Compton stage $\left( \nu_{m,c},\,S_{m,c}\right)$, the duration
of the Compton stage, $t_c$\footnote{Time is given throughout the
  paper in the observer's reference frame.}, and a given set of
exponents ({$s$, $b$, $k$, $d$}, and $\rho$). Some estimates for
the end of the Compton stage could be, for example, extracted from
blazar single-dish monitoring data of the F-GAMMA program.

Once the temporal evolution of the turnover frequency and the turnover
flux density is given, we can compute the single-dish light curves
using an approximation for the {SSA} spectrum:
\begin{equation}
S_\nu=S_\mathrm{m}\left(\frac{\nu}{\nu_\mathrm{m}}\right)^{\alpha_t}\frac{1-\exp{\left(-\tau_m\left(\nu/\nu_\mathrm{m}\right)^{\alpha_0-\alpha_\mathrm{t}}\right)}}{1-\exp{(-\tau_\mathrm{m})}},
\label{snu}
\end{equation}
where
$\tau_m\approx3/2\left[\left(1-8\alpha_0/{3\alpha_t}\right)^{1/2}-1\right]$
is the optical depth at the turnover frequency, $S_\mathrm{m}$ is the turnover
flux density, $\nu_\mathrm{m}$ is the turnover frequency, and $\alpha_t$ and
$\alpha_0$ are the spectral indices for the optically thick and
optically thin parts of the spectrum, respectively.

{The spectral shape is determined by two additional break frequencies, one at low frequencies, $\nu_\mathrm{l}$, and a high frequency break, $\nu_\mathrm{h}$. For frequencies lower than $\nu_\mathrm{l}$ the optically thick spectral index is given by $\alpha_\mathrm{t}=5/2$ whereas at higher frequencies the value for the optically thick spectral index is computed from the shock parameters of the adiabatic stage $\left(\alpha_\mathrm{t}=f_3/n_3\right)$. This low frequency break is motivated by the superposition of individual spectra with $\alpha_\mathrm{t}=5/2$ emitted from regions close to the shock front which leads to a flattening of the overall spectrum. Throughout the temporal evolution of the flare, $\nu_\mathrm{l}$ is set as a constant fraction, {$\zeta$} of the turnover frequency, $\nu_\mathrm{m}$ \citep[see][]{2000A&A...361..850T}}. {Since we are not fitting the model to the flare of an individual blazar, we assume $\zeta=0.6$ throughout the paper.}

 {At frequencies higher than $\nu_\mathrm{h}$ the  optically thin spectral index, $\alpha_0$ steepens by $-0.5$ due to synchrotron losses \citep[see, e.g.][]{1985ApJ...298..114M}. This break frequency depends on the energy of the relativistic electron distribution and on the magnetic field. Using the parameterisation of the energy and the magnetic field in terms of the jet radius, the evolution of the break frequency can be written as:}
\begin{equation}
\nu_\mathrm{h}\propto B\delta E^2\propto R^{-(4/3+b+d)}\propto t^\frac{-(4/3+b+d)\rho}{2d\rho+1}
\end{equation}
{Throughout the paper we assume that the break frequency is initially around $10^{12}$ -- $10^{13}$Hz and evolves according the above stated relation.}

{Following \citet{2000A&A...361..850T} we include an additional flattening of the optically thin spectral index, $\alpha_0$, from $-s/2$ to $-(s-1)/2$. This transition occurs usually during the transition from the synchrotron to the adiabatic stage and is modelled by:}
\begin{equation}
\alpha_0(t)=\frac{-s}{2}+\frac{1}{2}\frac{\log{t/t_\mathrm{trans}}}{\log{t_{2/3}/t_\mathrm{trans}}},
\end{equation}
{where $t_{2/3}$ is the transition time from the synchrotron to the adiabatic stage and $t_\mathrm{trans}$ is the time for the start of the spectral flattening, assuming $t_\mathrm{trans}<t_{2/3}$. Since $t_\mathrm{2/3}$ depends on the conditions in the jet, the transfer time is dynamically obtained in our model in order to obey the conditions above. See the top panel of Fig. \ref{vmsmplane} as an example for the influence of the break frequencies and the flattening of $\alpha_0$ on the spectral shape.}

\subsection{Shock-in-jet parameters}
It is {generally} assumed that the {flare} spectrum is
superimposed on top of a steady, quiescent spectrum
\citep[see][]{2011A&A...531A..95F}. We model this underlying spectrum
{using} a {SSA} spectrum with {an} optically thick
spectral index, $\alpha_t=5/2$. In Table~\ref{para} we present the
parameters used for both, the quiescent and the {flare} spectrum.
{To reduce the amount of free parameters in our analysis, we
  assume a conical jet ($\rho=1$) and a normalisation coefficient of
  the relativistic electron distribution decreasing adiabatically,
  which leads to $k=2(s+2)/3$.}

The remaining free parameters in Eqs.~3--8 are related to the
evolution of the magnetic field and the Doppler factor with distance,
given by $b$ and $d$, respectively (see
Eq.~\ref{an1} -- \ref{f3}), and the spectral slope of the relativistic
  electron distribution, $s$.  Parameter $b$ gives also information
about the structure of the magnetic field in the flow: if the magnetic
field is toroidal and if we assume an ideal plasma (implying that the
magnetic field is frozen to the particles), $b$ is equal to
1. Similarly, a poloidal magnetic field leads to $b=2$.
  Consequently, the exponent for the evolution of the magnetic field
varies between $1<b<2$. In the case of the spectral slope we study
values between {$2<s<3$} (and thus $2.7<k<3.3$), which gives the range of observed
spectral slopes.  Finally, the influence of {the Doppler factor}
$\delta$ on the spectral evolution is studied for values between
$d=-0.45$ to $d=0.45$. This range is motivated by the assumption that
the flux density should decrease with frequency in the Compton stage
($\epsilon_{\rm Compton} < 0$). Solving the expression for
$\epsilon_{\rm Compton}$ ($\epsilon_{\rm C}$, hereafter) for the
maximum allowed value of $d$ that gives $\epsilon_\mathrm{C}=0$, we obtain
$d=0.45$. {For the lower limit of $d$, we use a value of $d=-0.45$ leading to a symmetric parameter range for $d$. }{According to our definition (Eq.\,1), we thus
  investigate Doppler factors increasing along the jet for $-0.45<d<0$
  and decreasing for $0<d<0.45$, whereas $d=0$ denotes constant
  Doppler boosting along the jet.} The range for the exponents and the
values of the relevant parameters at the end of the Compton stage, as
{observed by the F-GAMMA program}, are summarised in Table~\ref{para}. 
{The latter are obtained from the variability and spectral
  analysis of the first 2.5 years of F-GAMMA monitoring data at 2.6 to
  228\,GHz demonstrating that maximum flux variations usually occur at
  mm wavelengths \citep{2014MNRAS.441.1899F}} {Based on the results of the F-GAMMA monitoring we compute the variability amplitude for all monitored sources. The peak of the variability amplitude is typically placed at frequencies between 43 GHz and 142GHz. Therefore, we use 86GHz as a reasonable estimate for the end of the Compton stage. 
 For the end of the Compton stage, $t_c$, we use the typical flare rise time (on-set of the flare to peak of the flare) at mm-wavelength which is roughly around 0.05\,yr. For most of the sources, the peak of the quiescent spectrum is out of the frequency range of the F-GAMMA monitoring. Therefore, we assume a low turnover frequency $\left(\nu_{m,q}=0.1\,\mathrm{GHz}\right)$ and turnover flux density of $S_{m,q}=5.4\,Jy$ and a spectral index, $\alpha_q=-0.5$. The reported values are first order estimates and could be improved by detailed modelling of the individual sources within the F-GAMMA monitoring which is out of the scope of this paper.}  

\begin{figure}[h!]
\resizebox{\hsize}{!}{\includegraphics{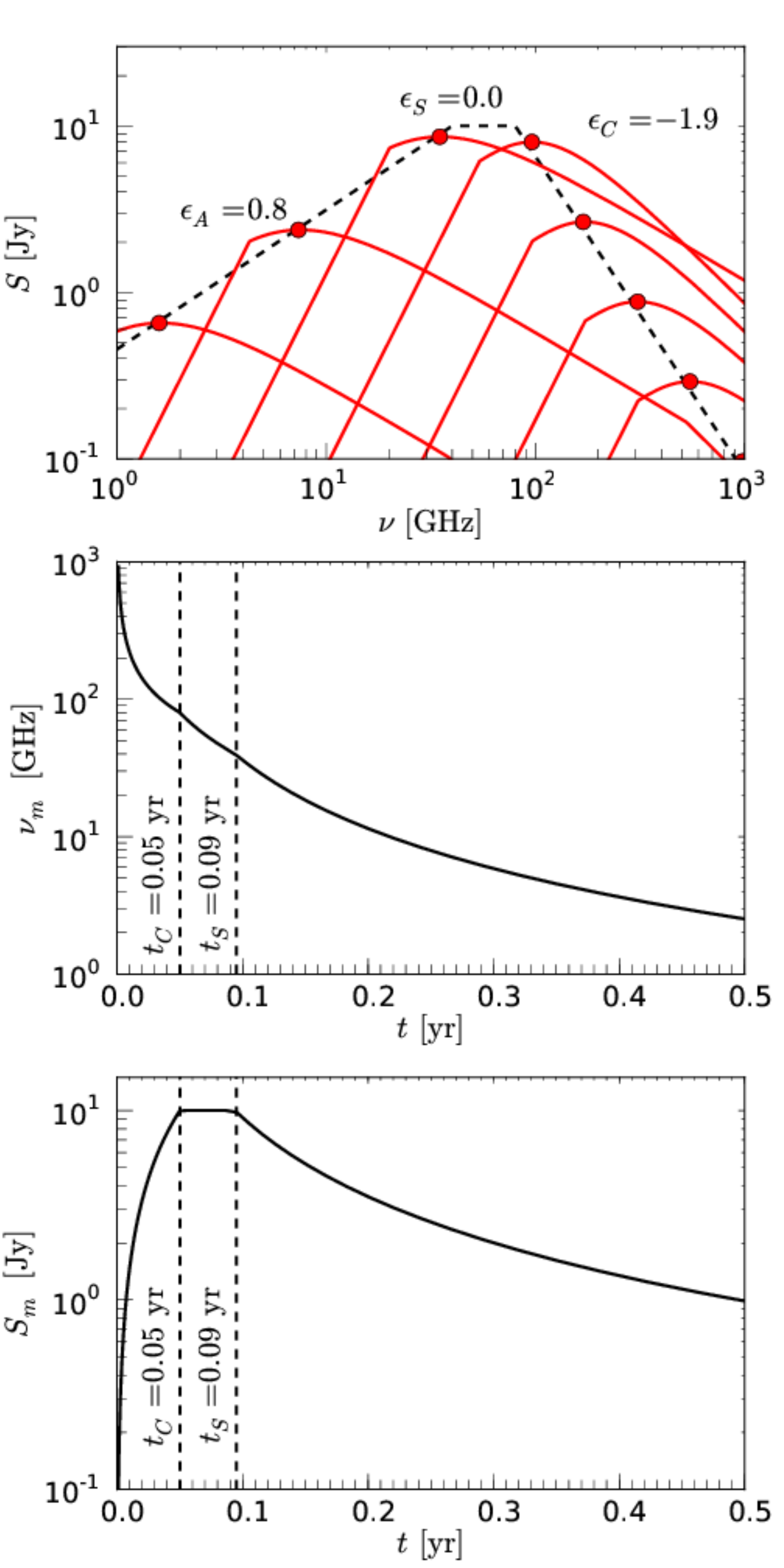}} 
\caption{Evolution of the turnover frequency, $\nu_\mathrm{m}$, and the
  turnover flux density, $S_\mathrm{m}$, for a model with $b=1.5$, $s=2.5$,
  $\rho=1$, $k=3.0$, and $d=0$. The top panel shows the evolution of the
  flare in the {$\nu_\mathrm{m}$--$S_\mathrm{m}$} plane (the time evolution is from
  the bottom right, to the centre and to the bottom left.). The black
  dashed line indicates the path along which {$\nu_\mathrm{m}$ and $S_\mathrm{m}$
    (red dots) evolve. The} slopes of the different stages are
  given in the plot. The red lines correspond to {SSA} spectra
  computed for different times (position along the spectral evolution
  path). The middle panel shows the temporal evolution of {
    $\nu_\mathrm{m}$}, and the bottom panel displays the temporal evolution of
  {$S_\mathrm{m}$}.}
\label{vmsmplane}
\end{figure}

In Fig.~\ref{vmsmplane} we present an example of a flare using
$d=0$ (constant Doppler factor), $b=1.5$ (helical field), $s=2.5$,
$k=3.0$ (fixed by $s$ following the assumption of adiabatic
expansion), and $\rho$=1.0 (conical jet).  The top panel { 
displays} the evolution of the flare in the turnover frequency
-- turnover flux density ($\nu_\mathrm{m}-S_\mathrm{m}$) plane. Some example {SSA}
spectra at different times during the flare are shown, including the low and high frequency spectral
breaks.

\begin{figure*}[t]
\centering 
\includegraphics[width=17cm]{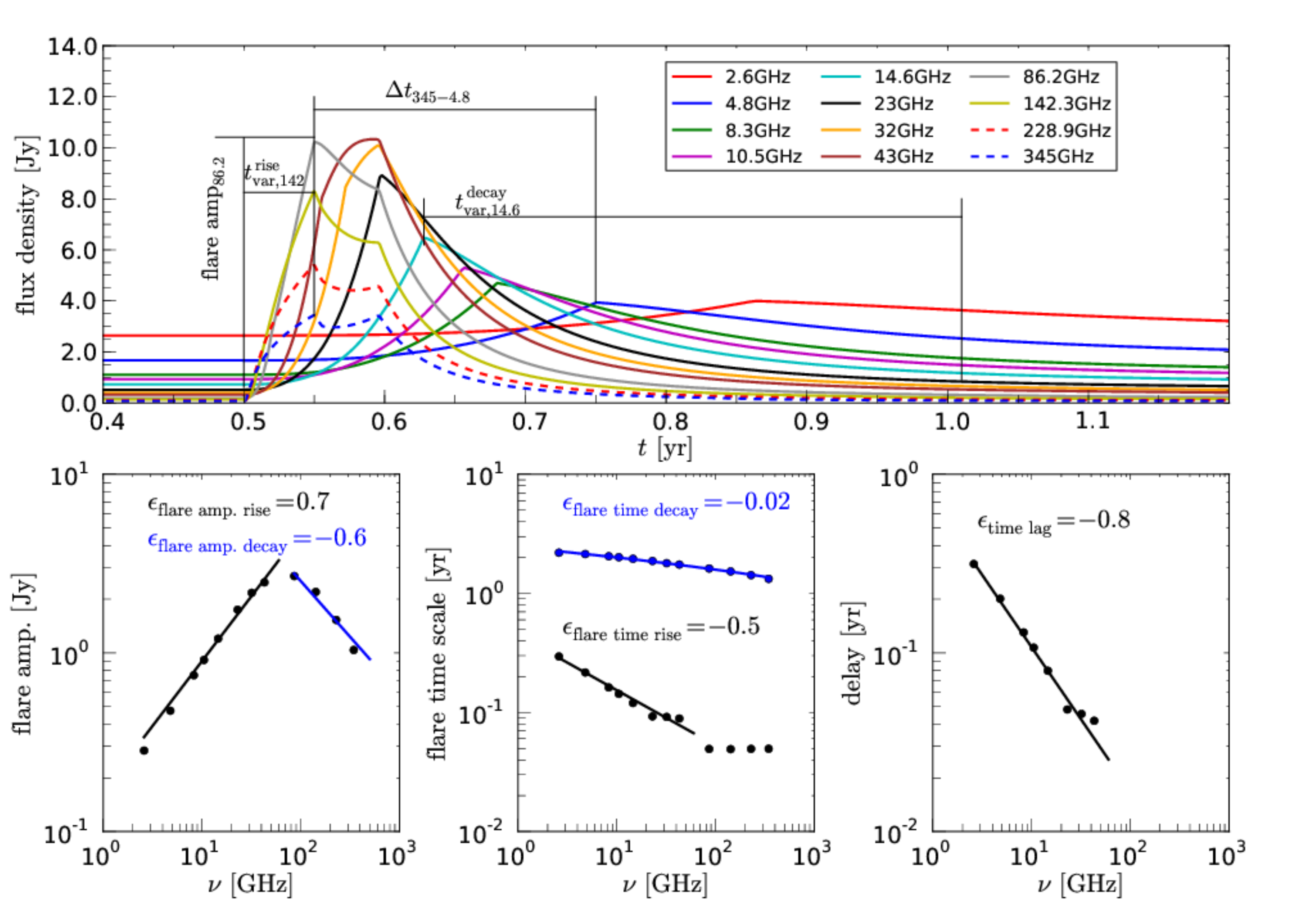}
\caption{Example for the analysis of the synthetic single-dish light
  curves. Top panel: the single-dish {multi-frequency light curves
    (2.6 to 345\,GHz)} computed for $b=1.5$, $s=2.5$, $k=3.0$, 
    $\rho=1.0$ and $d=0$ {(see also Fig.~\ref{vmsmplane})}.
  Bottom panel from left to right: {flare amplitudes, flare time
    scales} and cross-band delays. The solid lines correspond to
  power-law fits and the exponent is given in the plots.}
\label{para1} 
\end{figure*}

The middle and bottom panels
show the temporal evolution of the turnover frequency, $\nu_\mathrm{m}$, and
the turnover flux density, $S_\mathrm{m}$, respectively. The flare {
evolution along the three different energy loss stages can be best
seen in the $\nu_\mathrm{m}$--$S_\mathrm{m}$ plane (red circles along the dashed line
in Fig. 1, top)}. The flare starts at high turnover frequency and
low turnover flux density, and, as long as Compton losses are the
dominant energy loss mechanism, the turnover frequency decreases while
the flux density increases (see middle and bottom panels for
$t<0.05\,\mathrm{yr}$).
In the synchrotron stage ($0.05\,\mathrm{yr}<t<0.1\,\mathrm{yr}$) the
turnover flux density is constant (only for certain sets of
parameters, see Appendix) and the turnover frequency continues
decreasing, this leads to a plateau in the $\nu_\mathrm{m}$--$S_\mathrm{m}$ plane. In
the final stage of the shock evolution, the main losses are due to the
adiabatic expansion of the jet and both, turnover frequency and
turnover flux density decrease ($t>0.1\,\mathrm{yr}$).

\begin{table*}[t!]
\caption{Parameters used for the modelling of the spectral evolution.}
\label{para}
\centering
\begin{tabular}{@{}c c c c c c c c c c@{}}
\hline \hline
$b$ & $s$ & $k$ & $d$ &$\rho$ & $t_c$ & $\nu_\mathrm{m,c}$ & $S_\mathrm{m,c}$ & $S_{q,\mathrm{c}}$ & $\alpha_q$ \\
$[1]$ & [1] & [1] & [1] &[1]  & [yr] & [GHz] & [Jy]  & [Jy] & [1]\\
\hline
1--2 & 2--3 & 2.7--3.3$^\star$  &$-$0.45--0.45$^\ast$& 1 & 0.05$^\dagger$ & 80$^\dagger$ & 10$^\dagger$ & 5.4 & $-$0.5\\
\hline
\multicolumn{9}{l}{$^\ast$ required to ensure that $\epsilon_{\rm C}<0$, $^\star$ assuming adiabatic losses $k=2(s+2)/3$}\\
\multicolumn{9}{l}{$^\dagger$ fixed value {as observed by the F-GAMMA program}}\\
\end{tabular}
\end{table*}

%
% In the bottom panels of Fig. 2 the labels for the parameters should
% be the same as those used in the text/definitions,
% i.e. $\sigma_{s}$, $\Delta t$. For the delay I suggest to generally
% use $\tau$
%

\subsection{Synthetic light curve analysis}\label{slca} 
We produced synthetic total flux density light curves at different
radio frequencies by sweeping through the parameter space of exponents
$b$, $s$, and $d$, as given in Table~\ref{para}. These light curves
are similar to those typically acquired by multi-frequency single-dish
blazar monitoring programs (the F-GAMMA program for instance). {In
total we calculated light curves for 12 ``standard'' radio frequency
bands, namely 2.6\,GHz, 4.8\,GHz, 8.3\,GHz, 10.5\,GHz, 14.6\,GHz, 23\,GHz, 32,GHz, 43\,GHz, 86.2\,GHz, 142.3\,GHz, 228.9\,GHz, and 345\,GHz}. In addition, we extracted a
series of {multi-frequency} parameters {from each set of light
  curves} that could be compared to single-dish observations. Our aim
was to produce a catalog of spectral evolution types that could be
used as patterns to derive jet parameters from observed light
curves. The parameters that we propose are the following:

\begin{itemize}
\item {\bf Flare amplitudes}: In order to characterise the
  frequency-dependent strength of the synthetic flares, we compute the
  light curve standard deviations $\sigma_{s}$ from the ground values
  in Fig.~\ref{para1}  and apply a power-law fit with frequency
  $\sigma_{(S)}\propto\nu^{\epsilon_{\mathrm{flare\,amp.}}}$.
  {Since the flare amplitudes show a rising and a decaying part, we apply
  a power law fit to each part separately}

\item {\bf Flare time scales}: {The time scale of a flare can be obtained from the rising 
and/or from the decaying edge of the individual light curves.
The rising time scale of a given flare is  obtained as $\Delta t=t_\mathrm{max}-t_\mathrm{min,r}$ 
(see also Fig.~\ref{para1}), with the time between the start of flux density increase ($t_\mathrm{min,r}$)
  and the time at the flare maximum ($t_\mathrm{max}$). Whereas the decaying time scale is computed 
  as $\Delta t=t_\mathrm{min,d}-t_\mathrm{max}$, with $t_\mathrm{min,d}>t_\mathrm{max}$ the time required to obtain quiescent flux density (see Fig.~\ref{para1}). We extract these two time scales from the light curves and fit the obtained data with
  a power-law, $\Delta t\propto\nu^{\epsilon_\mathrm{flare\,time}}$.}

%\label{vartimescale} 

\item {\bf Cross-band delays}: In general, there is a delay between
  the flux density peaks at different frequencies. In order to
  quantify these multi-frequency {delays}, we compute the time
  difference between the flux density peak at our highest frequency
  (345\,GHz) and the flux density peaks of the other frequencies. 
  The growth of the cross-band delays with decreasing
  frequency was approximated with a power-law fit
  $\left(\tau_{345-\nu_i}\propto\nu^{\epsilon_{\mathrm{delay}}}\right)$.

\end{itemize}

These frequency-dependent single-dish {light curve} parameters
will be referred to as {light curve} parameters from now on.

\subsection{An example}
{In Fig.~\ref{para1} we show the analysis of the synthetic light curves
computed for the set of parameters also used in Fig.~\ref{vmsmplane}: 
$b=1.5$, $s=2.5$, $k=3.0$, $\rho=1.0$ and $d=0$. While Fig.~\ref{vmsmplane} 
shows the evolution of $\nu_\mathrm{m}$ and $S_\mathrm{m}$ as well as example spectra during 
the evolution of the flare, the upper panel of Fig.~\ref{para1} presents 
the computed single-dish light curves from 2.6\,GHz to 345\,GHz.} 
In this panel, the measure of the {light curve} parameters ({flare} 
amplitudes, time scales and cross-band delays) is
indicated. The high-frequency light curves ($\nu\ge 86\,\mathrm{GHz}$)
show a rapid rise and reach their peak flux density nearly
simultaneously. The increase in the time delay between the highest
frequency (here 345\,GHz) and the lower frequency light curves
($\nu<86\,\mathrm{GHz}$) is clearly visible. {The amplitude of the flux
density peak first increases towards lower frequencies, peaks around 
$\nu=86\,\mathrm{GHz}$, and then decreases, while} the time
required for obtaining the flux density peak at each frequency
increases with decreasing frequency (best visible at 2.6\,GHz). The
analysis of these trends can be found in the bottom panels { of Fig.~\ref{para1}}, which
show, from left to right, the {flare amplitudes $\sigma_{s}$, time
  scales $\Delta t$, and cross-band delays $\Delta t_{345-\nu_i}$}
versus frequency. The solid line in each of those panels corresponds
to a power-law fit {(see Sect. \ref{slca})} and the exponents are
given in the plots. {Notice that for frequencies higher than 43\,GHz no delays are obtained for this example.}

\section{Results}
In this Section we present the results for the variation of the slopes
of the different energy loss stages in the $\nu_\mathrm{m} - S_\mathrm{m}$~plane, and
{the corresponding changes} of the light curve parameters, as a
function of $b$, $s$, and $d$ (we recall that $\rho=1$ and {
  $k=2(s+2)/3$}). Since the range of possibilities given by the
different combinations of the parameters is large, we only show the
results for selected models where we fixed either $b=1.5$ (helical
magnetic field) or $s=2.5$ (a typical value of a spectral slope), while
varying parameter $d$, which controls the evolution of {the}
Doppler factor. This is motivated by the results being most sensitive
to changes in this parameter, as we show {below}. The results for the
entire parameter space can be found in the Appendix.

\begin{figure}[h!]
\resizebox{\hsize}{!}{\includegraphics{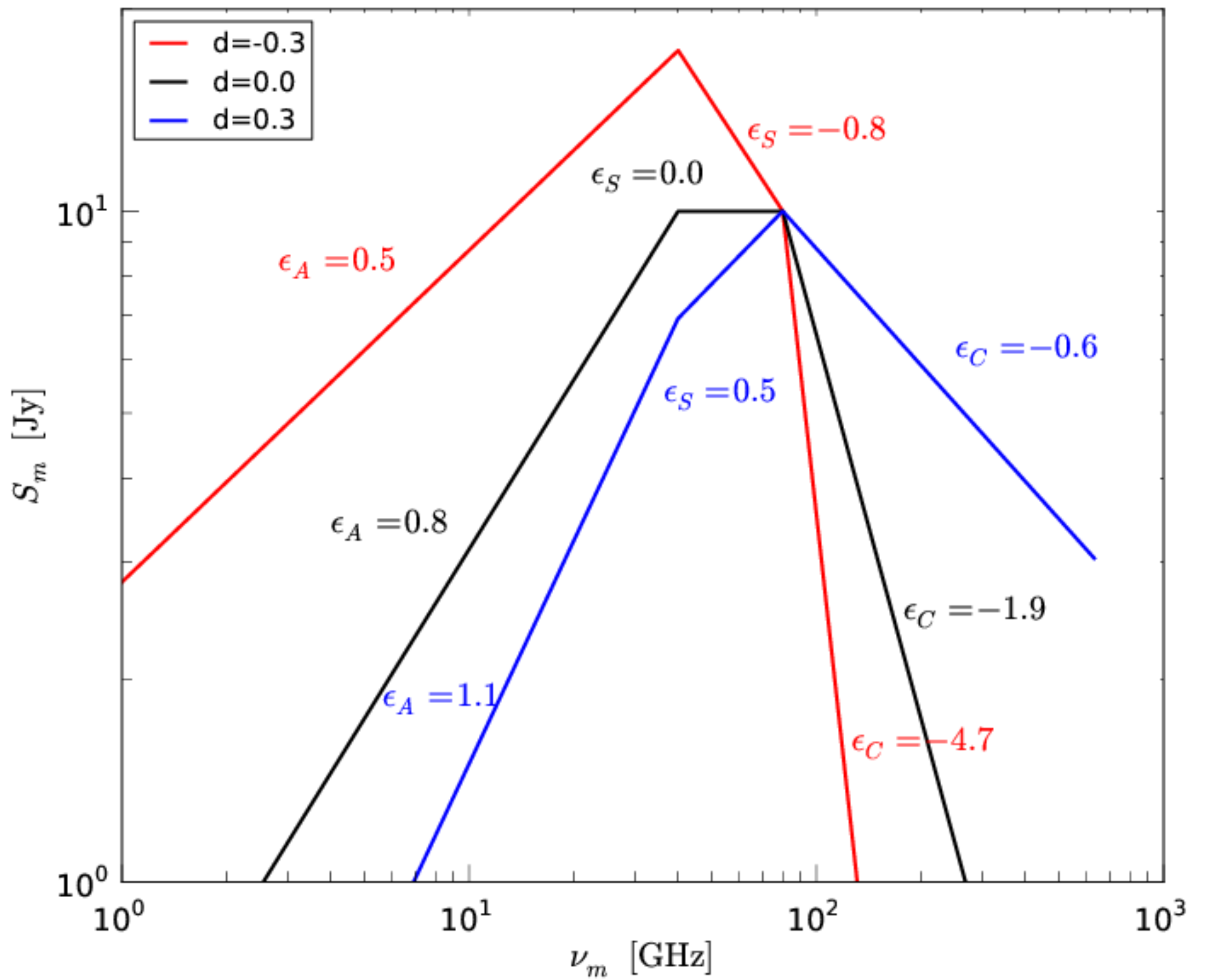}} 
\caption{Evolution of the flare in the turnover frequency -- turnover
  flux density plane for three different values of $d$ while keeping
  $b=1.5$, $s=2.5$, $\rho=1$, and, $k=3.0$ fixed. The slopes of the
  different stages are given in the plot and are color-coded.}
\label{vmsmplane1}
\end{figure}

\begin{figure*}[t]
\centering 
\includegraphics[width=17cm]{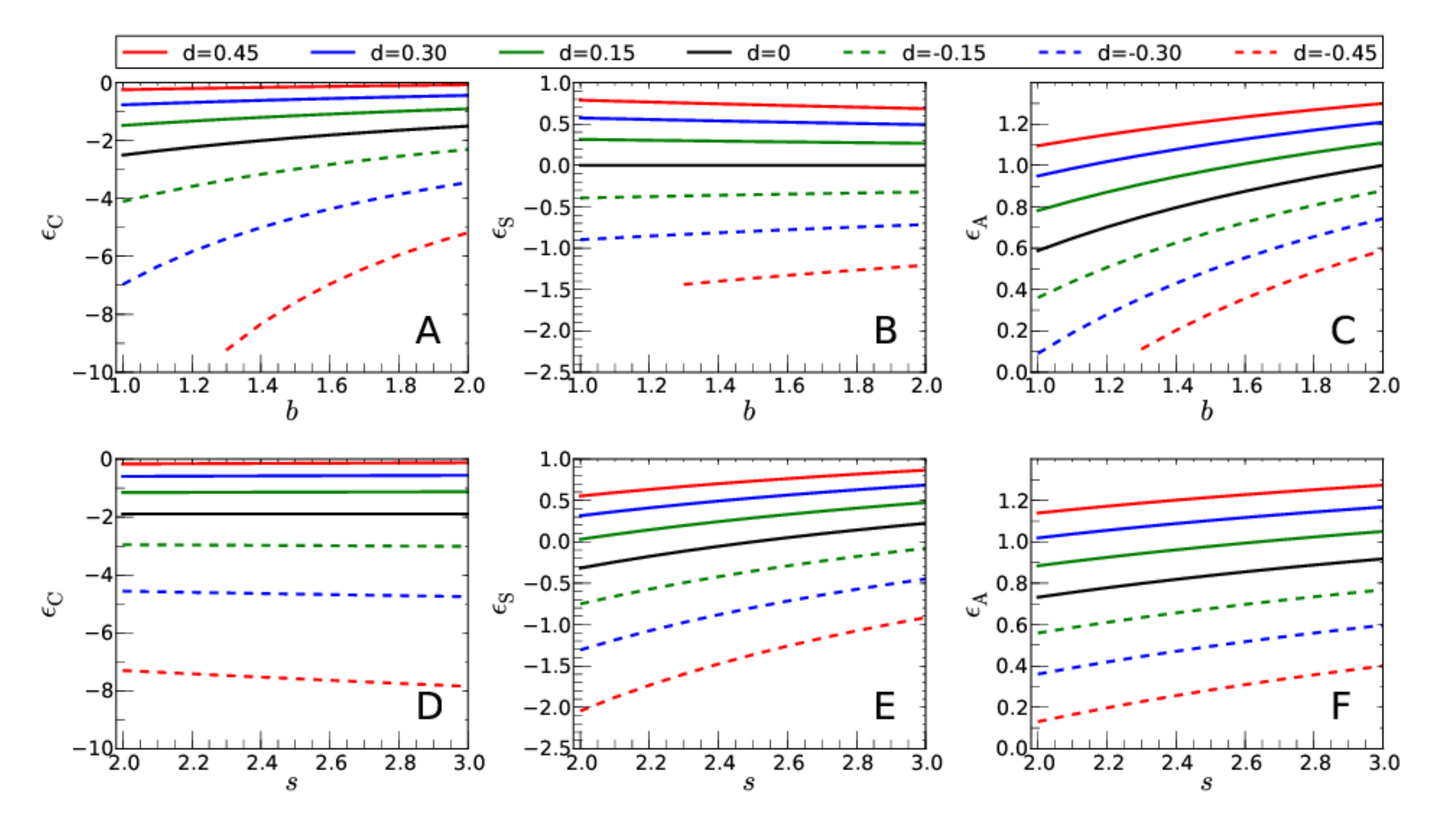}
\caption{Exponents for the different energy loss stages from left to
  right: Compton (A,D), synchrotron (B, E) and adiabatic stage
  (C,F). The top panels (A--C) show the variation of the slopes using
  a fixed spectral slope of $s$=2.5 and the bottom panels (D--F)
  present the changes in the slopes for a fixed helical magnetic field
  ($b$=1.5). The different values for $d$ are color-coded and can be
  found in the plot legend.}
\label{paracut1} 
\end{figure*}

\subsection{Slopes of the energy loss stages}
The results of our modelling show that the evolution of the Doppler
factor along the jet, which is controlled by parameter $d$, has the
largest impact on the variation of the slopes for the different energy
loss mechanisms in the $\nu_\mathrm{m} - S_\mathrm{m}$~plane (see the
Appendix). Figure~\ref{vmsmplane1} shows the changes in the slopes
that characterise the different stages, for three values of $d$
($d=-0.3$, $d=0$, and, $d=0.3$) while keeping $b=1.5$, $s=2.5$, and
$k=3.0$ fixed. We remind the reader that the definition of $d$ in
Eq.~\ref{eq:bkr} results into an increasing Doppler factor with distance for $d<0$
and a decreasing Doppler factor for $d>0$. The slopes of
the Compton ($\epsilon_{\rm C}$) and the adiabatic stage
($\epsilon_{\rm Adiabatic}$, $\epsilon_{\rm A}$ hereafter) increase
with $d$, { i.e. the Compton stage flattens while the
  adiabatic stage steepens for increasing values of $d$} (see
Fig.~\ref{vmsmplane1}). Interestingly, there is a change in the sign
of the synchrotron stage ($\epsilon_{\rm Synchrotron}$, $\epsilon_{\rm
  S}$ hereafter): the slope is positive for $d>0$ and negative for
$d<0$. This implies that the difference between the slopes of the
Compton and the synchrotron stage decreases for $d<0$ (red line in
Fig.~\ref{vmsmplane1}), which can cause an identification of the
latter within the former, while for $d>0$ the same happens with the
slopes of the synchrotron and the adiabatic stages (blue line in
Fig.~\ref{vmsmplane1}). {Consequently}, the synchrotron stage
could be difficult to identify in single-dish observations. Another
important conclusion is that, assuming that the viewing angle to the
flaring flow does not change during this time, acceleration of the
flow would produce a synchrotron stage with negative slope, whereas a
decelerating flow would result into a positive slope.

\begin{figure}
\resizebox{\hsize}{!}{\includegraphics{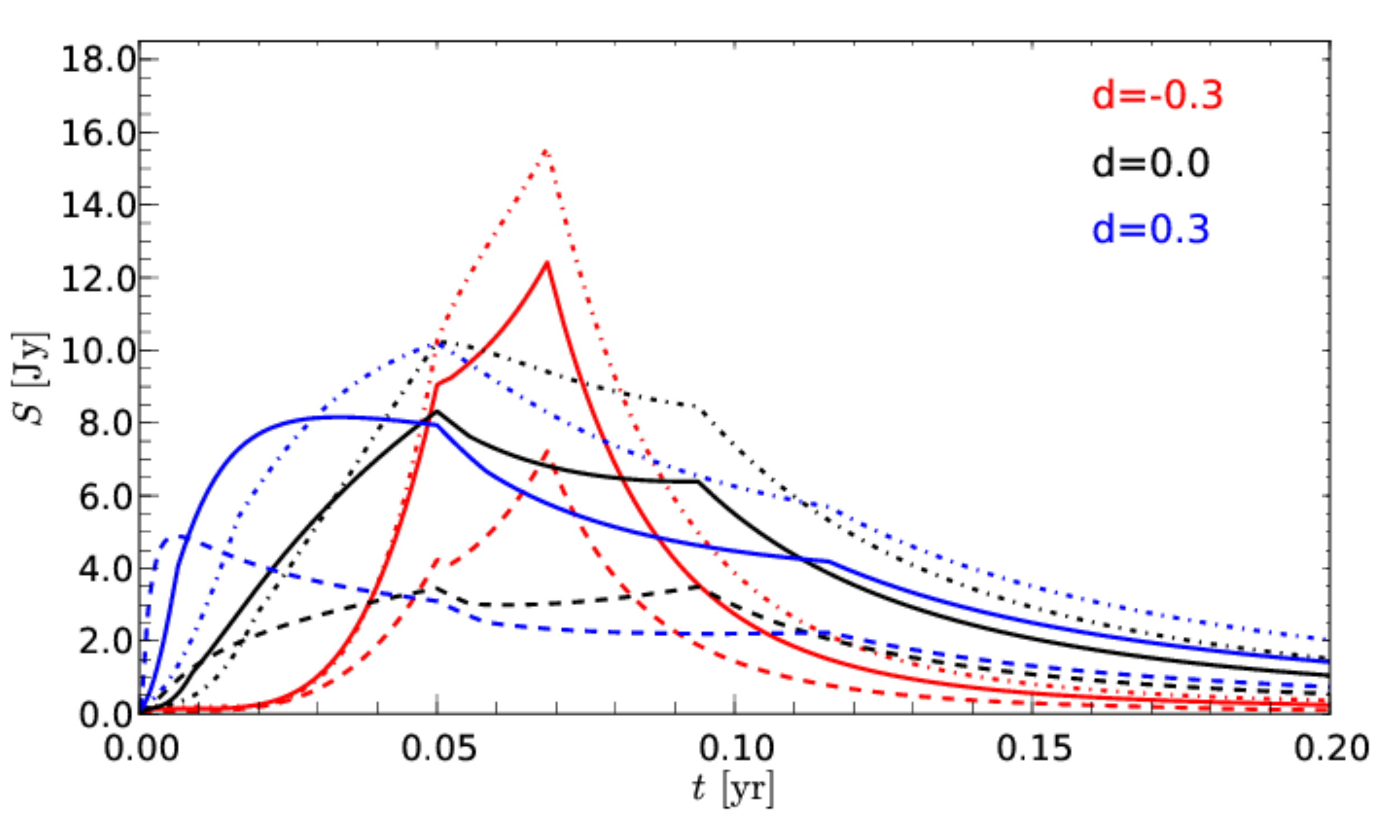}} 
\caption{High-frequency {(86 to 345\,GHz)} light curves computed
  for three different values of $d$ while keeping $b=1.5$, $s=2.5$,
  $\rho=1$, and, $k=3.0$ fixed. The dashed lines correspond to
  345\,GHz, the solid ones to 140\,GHz and the dot-dashed ones to
  86\,GHz. {See text for more details}.}
\label{lightcurves}
\end{figure}

The dependence of the slopes of the different energy loss stages in
the $\nu_\mathrm{m} - S_\mathrm{m}$~plane with parameters $b$ and $s$, for different
values of $d$ {($-0.45<d<0.45$),} is presented in
Fig.~\ref{paracut1}. The upper panels show the variation of the
slopes with $b$ {($1<b<2$)}, taking $s=2.5$, whereas the bottom
panels show the variation of the slopes with $s$ {($2<s<3$)},
taking $b=1.5$. The results obtained can be summarised as follows:

\paragraph{Compton stage ($\epsilon_{\rm C}$):}
The slope of the Compton stage is not sensitive to changes in $b$ for
values larger than {about} 1.5 (panel A in Fig.~\ref{paracut1}) or
positive values of $d$, { while for (large) negative values of $d$
the Compton stage steepens strongly for values $b\lesssim 1.5$. 
$\epsilon_{\rm C}$} also shows little sensitivity with respect to the 
spectral slope $s$ (panel D in Fig.~\ref{paracut1}).

\paragraph{Synchrotron stage ($\epsilon_{\rm S}$):}
While $b$ increases, the slope of the synchrotron stage decreases for
jets with $d$>0 and increases for jets with $d$<0 (panel B in
Fig.~\ref{paracut1}). The slope of this stage increases with
increasing $s$ (panel E in Fig.~\ref{paracut1}).

\paragraph{Adiabatic stage ($\epsilon_{\rm A}$):}
The slope of the final stage increases with $b$ and $s$, being
slightly more sensitive to the former (see panel C and F in
Fig.~\ref{paracut1}).

\begin{figure*}
\includegraphics[width=17cm]{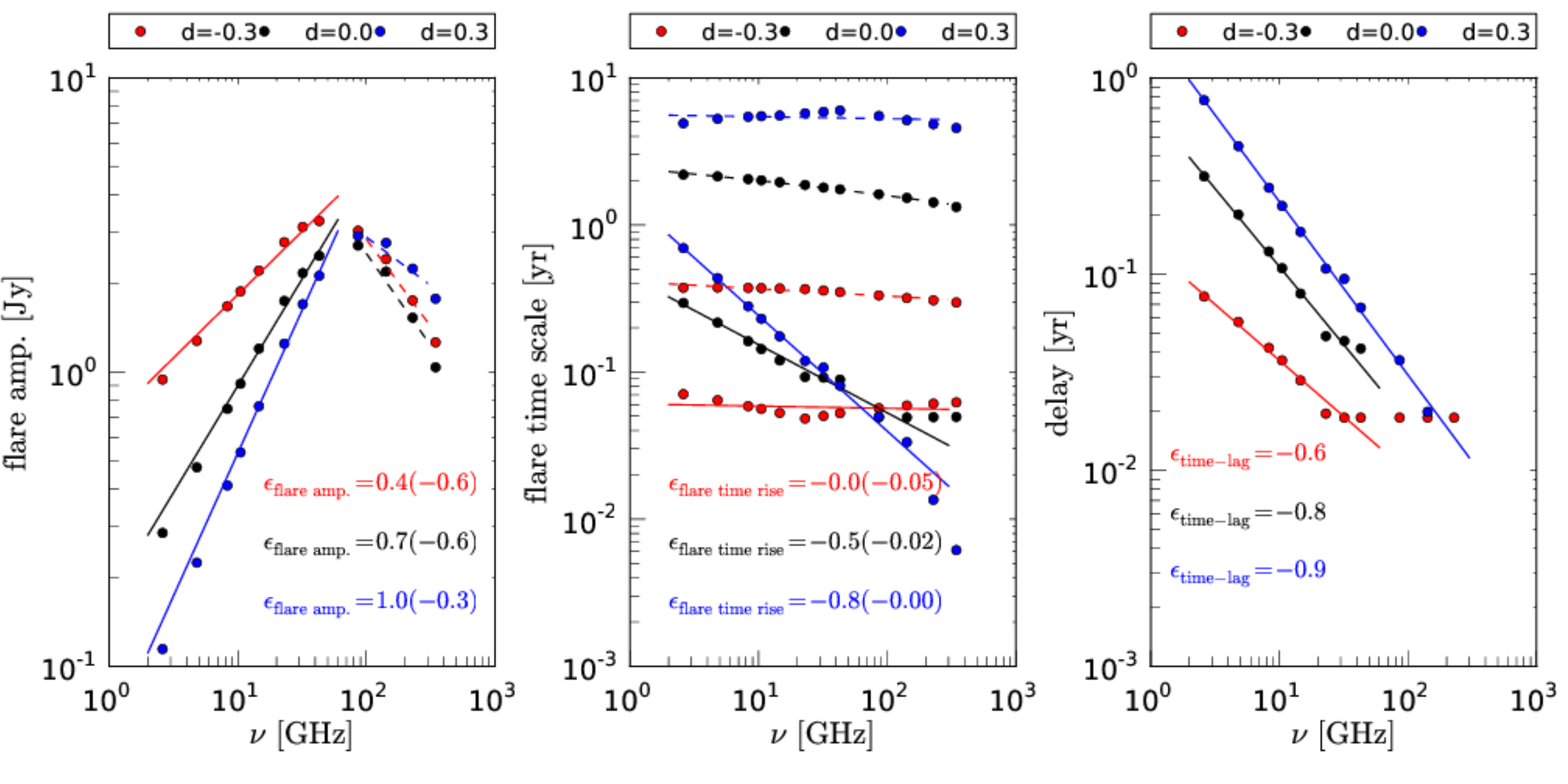}
\caption{Light curve parameters obtained from synthetic light curves
  with three different values of $d$ while keeping $b=1.5$, $s=2.5$,
  $\rho=1$, and, $k=3.0$ fixed. The panels show from left to right the
  flare amplitude, the flare time scale and the cross-band delays
  with respect to the peak of the 345\,GHz light curve. The solid
  lines correspond to the power-law fits and the exponents are given
  in the plots. {The values in brackets correspond to the exponents of a power law fit to the 
  decaying regime of flare amplitude and flare time scale.}}
\label{freqpara}
\end{figure*}

\begin{figure*}
\centering 
\includegraphics[width=17cm]{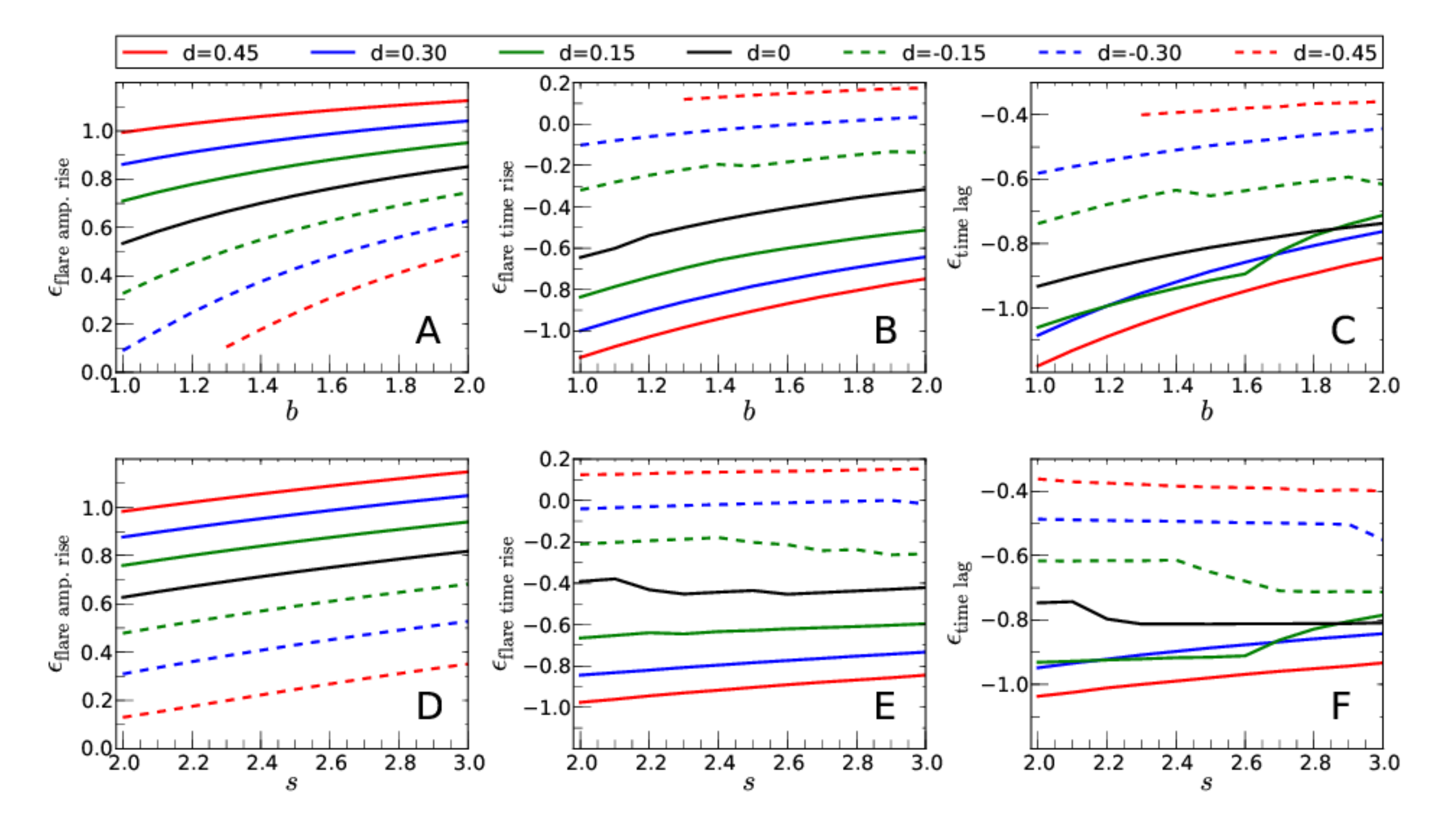}
\caption{Exponents of the light curve parameters obtained from the
  synthetic light curves from left to right: {flare}
  amplitude, {flare} time scale and cross band delay exponent. The top panels show
  the variation in the light curve parameters using a fixed spectral
  slope of $s$=2.5 (panels A--C) and the bottom panels (D--F) present the changes in the
  parameters for a fixed helical magnetic field ($b$=1.5). The
  different values for $d$ are color-coded and can be found in the
  plot legend.}
\label{paracut2} 
\end{figure*}

\begin{figure*}
\centering 
\includegraphics[width=17cm]{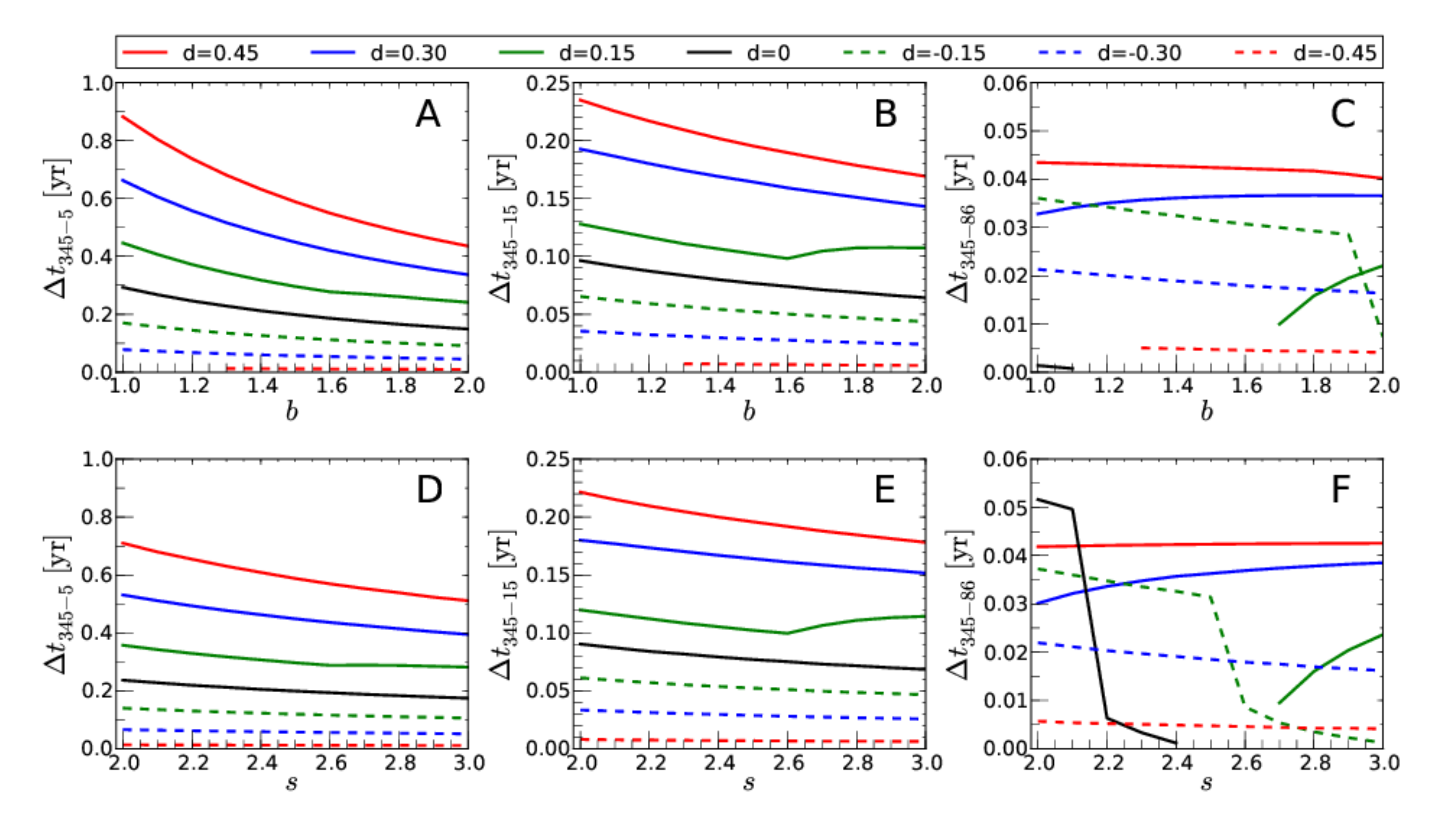}
\caption{Cross-band delays between 345\,GHz and three selected frequencies (from left to right 5~GHz, 15~GHz and 86~GHz).
  The top panels (A--C) show the variation in the cross-band delays using a fixed spectral slope of $s$=2.5 and
  the bottom panels (D--F) present the changes in the cross-band delays for a
  fixed helical magnetic field ($b$=1.5). The different values used
  for $d$ are color-coded and can be found in the plot legend.}
\label{paracut3} 
\end{figure*}

\subsection{Light curve parameters}
To better understand the results of {our analysis with respect to
  the} synthetic single-dish light curves it is necessary to
investigate the influence of the shock parameters on the shape of the
light curves, especially above 86\,GHz. {In Fig.~\ref{lightcurves}
we present light curves at 345\,GHz (dashed lines), 140\,GHz (solid
lines), and 86\,GHz (dot-dashed lines) for $d$=0 (black lines),
$d$=0.3 (blue lines) and $d$=$-$0.3 (red lines) while keeping 
$b=1.5$, $s=2.5$ and $k=3.0$}. For {$d=0$} the
shape of these high-frequency light curves is basically the same and
the {flare peak} is reached simultaneously (black lines). For
$d>0$, a time delay between the peaks in the light curves is
observed. {For $d\geq0$, the shape of the curves changes with frequency:
the light curves steepen after the peak as the frequency is decreasing
(black and blue lines)}. In the case of $d<0$, {the 86 and 142\,GHz flares
  peak at the same time but delayed w.r.t. the reference frequency
  (345\,GHz), i.e. a constant delay between the peaks at
  $\nu=86\,\mathrm{GHz}$ and $\nu=140\,\mathrm{GHz}$ and 345\,GHz is
  observed}. This behaviour can be explained by the rising turnover
flux density during the synchrotron stage  (see
  Fig.~\ref{vmsmplane1}), which brings the peaks at high-frequencies
(those to the right of the one giving the maximum flux-density of the
flare) to occur at the same time, and the cut-off frequency, which
limits the peak flux-density at 345\,GHz to the time when the cut-off
crosses it (see Fig.~\ref{vmsmplane}).

Parameter $d$ has the largest impact on the light curve parameters,
too. As an example, we show in Fig.~\ref{freqpara} the computed light
curve parameters for the three different values of $d$ while fixing
$b=1.5$, $s=2.5$, $\rho=1$, and $k=3.0$. The panels show, from left to
right, the {flare} amplitude, flare time scale and the cross-band delay relative
to the peak of the 345\,GHz light curve. The solid and dashed lines in each panel
correspond to power-law fits as described in Sect.~\ref{slca}. {The slope of the flare amplitude for the rising and decaying edge (left panel
in~Fig.~\ref{freqpara}), given by the exponent $\epsilon_\mathrm{flare\,amp.}$, increases with $d$ (The values in brackets correspond to the decaying edge).}
For $d<0$, the {flare} time-scale (middle panel in
Fig.~\ref{freqpara}) approaches a constant value for all frequencies, as opposed to $d>0$, in which case it
continues decreasing with frequency (blue dots in the
panel). This translates into a decrease of the {flare} time scale
exponent with $d$, which implies that the {flare} time scale is
steepening with $d$. Only for $d>0$ we find an increasing time delay
between all frequencies (see right panel in Fig.~\ref{freqpara}). In
the case of $d<0$, we observe a constant delay among frequencies
$\nu>86\,\mathrm{GHz}$ and an increase for smaller frequencies. If
$d=0$, there is no delay between the frequencies
$\nu>86\,\mathrm{GHz}$. The exponent obtained from a power-law fit to
the time-lags decreases with {increasing $d$}.
     
The influence of $b$ and $s$ on the frequency-dependent light curve
parameters are presented in Fig.~\ref{paracut2}. 
As in Fig.~\ref{paracut1}, the upper panels show the variation of the slopes
with $b$, taking $s=2.5$, whereas the bottom panels show the variation
of the slopes with $s$, taking $b=1.5$. The results obtained can be
summarised as follows:

\paragraph{The {flare} amplitude exponent $\left(\epsilon_\mathrm{var.\,amp}\right)$:}
The values for the flare amplitude exponent increase
while the magnetic field configuration is changing from toroidal ($b=1$) to poloidal ($b=2$)
and with $s$. The variation in $\epsilon_\mathrm{var.\,amp}$ is slightly larger with $b$ than with $s$ (see panel E of Fig.~\ref{paracut2}). The influence of $b$ and of $s$ on $\epsilon_\mathrm{var.\,amp}$ decreases with $d$ (see panel A and E in Fig.~\ref{paracut2}).

\paragraph{The {flare} time scale exponent $\left(\epsilon_\mathrm{var.\,time\,scale}\right)$:}
Exponent $\epsilon_\mathrm{var.\,time\,scale}$ increases while $b$
changes from b=1 (toroidal field) to b=2 (poloidal field) (panel C of
Fig. \ref{paracut2}). Finally, we observe only a minor variation of
$\epsilon_\mathrm{var.\,time\,scale}$ with increasing $s$ for $d<0$,
and an increase when $d>0$ (panel F of Fig. \ref{paracut2}).

\paragraph{The cross-band delay exponent $\left(\epsilon_\mathrm{delay}\right)$:}
The distribution of values for the delay exponent strongly depends
on the value of $d$. The exponent is strongly increasing with $b$ for $d>0$ and only a minor increase is obtained for $d<0$ (see panel C of Fig~\ref{paracut2}). {The delay exponent increases with $s$, except for $d=0$ and $d=-0.15$, where the delay exponent decreases with increasing $s$. The variation in the cross-band delay exponent is smaller than for $b$ (see panel F of Fig~\ref{paracut2} and decrease with $d$)}

\paragraph{Time lags with respect to 345\,GHz:}
In Fig.~\ref{paracut3} we present time lags for the peak fluxes at
5\,GHz, 15\,GHz, and 86\,GHz, with respect to 345\,GHz. The top
panels show the variation of the time lag for a constant spectral
slope, $s$=2.5, while changing $b$ from 1 to 2, whereas the bottom
panels show these variations for $b=1.5$ (helical magnetic field), and
spectral slope ranging from $s$=2 to $s$=3. The general trend for
$\nu<86$\,GHz can be described as follows: The time lags increase with
increasing $d$, as expected from Fig.~\ref{freqpara}. Regarding the
influence of $b$, the time lags significantly decrease with increasing
$b$. Finally, there is only a minor decrease in the time lags with
increasing $s$. {The obtained time lags at high frequencies, panels C and F, are very small $<18$ days. The general trend for the variation of the time lags can be summarised as follows: For $d>0$ the time lags are nearly constant and decrease with $s$ and $b$ for $d<0$.}

\subsection{The effects of second order Compton scattering}
\cite{2000ApJ...533..787B} re-evaluated the spectral evolution within
{the} Compton stage of the classical shock-in-jet model of
\cite{1985ApJ...298..114M} by including second order Compton
scattering and showed that {this} leads to flatter {slopes} in
the $\nu_\mathrm{m}$--$S_\mathrm{m}$ plane. {Within our model this is obtained by replacing $n_1$ and $f_1$, as given by Eq. 3 and Eq. 6 with:}
\begin{eqnarray}
n_{1,\mathrm{mod}}&=&-(4s+8+3bs+12b)/[3(s+12)]\\
f_{1,\mathrm{mod}}&=&[-4(s-13)-6b(s+2)]/[3(s+12)].
\end{eqnarray}

The Compton stage for $d=0$ and the results for the slopes are presented
in the first row of Fig.~\ref{modcompton}. The range of the slope
varies between $-0.6<\epsilon_{\rm C}<0.3$. The positive value of
$\epsilon_{\rm C}$ indicates an increase of the flux density with
frequency in the Compton stage, which is opposed to the observed
evolution during flares. We thus find that this model is only valid
for a limited range of shock parameters ($1<b<1.4$ and $2<s<3$) (see
Fig.~\ref{modcompton}). In contrast to the standard shock-in-jet model
with $d=0$, this model results into non-zero time delays between the
highest frequencies (see right panel in the third row in
Fig.~\ref{modcompton}).

\section{Discussion}
\subsection{Emission, absorption and expansion}
Overall, the different exponents and light curve parameters depend on
the delicate interplay between emission and absorption that controls
the spectral evolution of the flares. To discuss this we will restrict
ourselves to Figs.~\ref{vmsmplane1} to \ref{paracut3}, as the
argumentation remains valid for the global picture drawn by
Figs.~\ref{paraspace1} to \ref{paraspace6} (see the Appendix). During
the Compton stage the main energy-loss mechanism is inverse Compton
scattering. The radio spectrum is thus conformed {to} the synchrotron
emission and the amount of high-energy electrons/positrons available
to up-scatter the photons. The slope {$\epsilon_{\rm C}$ of this stage}
depends on the delay of the passage of the spectral peak through the
different frequencies. The smaller the delay, the steeper the
slope. The radio emission rises within this stage due to a decrease in
the opacity owing to a drop of the amount of available high-energy
electrons/positrons. 
{For $d<0$ (increasing Doppler factor with distance) the slope
  becomes steeper, i.e. the rise in $S_\mathrm{m}$ faster: assuming a constant
  viewing angle, the acceleration brings the flare to faster
expansion and cooling and a consequent loss of high-energy pairs,
whereas if the velocity is constant, a decreasing viewing angle
implies an increase in the observed emission 
{while} the flow expands and the loss of high-energy pairs also occurs,
although at a slower rate. Both possibilities make $\epsilon_{\rm C}$ steeper.
The exponent that controls the evolution of the magnetic field
intensity has a role, although secondary, in the value of
$\epsilon_{\rm C}$. {The} smaller the parameter $b$ is, the slower 
is the decrease of the intensity and the higher the synchrotron emissivity
remains along the evolution of the flare. {This} gives a fast increase
in $S_\mathrm{m}$ as the density of absorbing high-energy particles falls with
distance. The spectral index does not play a significant role, other
than increasing the relative amount of available high-energy particles
for smaller values of $s$. {This}, in addition, decreases the
emissivity at {lower} frequencies, thus having a negligible overall
effect on the development of this stage.

During the synchrotron {and adiabatic stages}, the spectral peak
is basically controlled by synchrotron self-absorption. An accelerated
expansion or a decrease of the viewing angle produces a further
increase in $S_\mathrm{m}$ while the peak frequency decreases within the
synchrotron stage, owing to the faster drop in non-thermal particle
density (i.e., a drop in the absorption coefficient) or larger
boosting, respectively, while the intrinsic emissivity at those
frequencies is still high. Flow deceleration or an increase of the
viewing angle would have the contrary effect. Only in the case of
$d=0$ we obtain a flat slope in the $\nu_\mathrm{m} - S_\mathrm{m}$~plane. The electron
distribution $s$ plays a more important role than during the Compton
stage, with flatter electron distributions reducing the steepness of
positive slopes ($\epsilon_{\rm S}>0$) and increasing the steepness of
negative slopes ($\epsilon_{\rm S}<0$). In the former case, a flatter
distribution of electrons results into a more important loss of
absorbing electrons at the highest energy end of the spectrum (they
cool-down faster), implying a faster drop in the peak frequency
relative to that in the peak flux, thus giving a flatter
$\epsilon_{\rm S}$. In the latter, the same loss of absorbing
electrons implies a continued increase of received flux and, thus, a
steeper negative slope. The evolution of the magnetic field is less
important to this stage because, on the one hand, a faster drop in the
field implies a decrease in the emissivity but, on the other hand,
also {a} smaller absorption coefficient. The adiabatic stage shows
a very similar trend relative to the relevant parameters. However,
$\epsilon_{\rm A}$ is more sensitive to the evolution of the magnetic
field intensity: smaller values of $b$ imply a smaller change in $S_\mathrm{m}$
relative to the drop in $\nu_\mathrm{m}$, because the emissivity decrease with
distance is slower than for {larger} values of $b$.
      
%new paragraph version of Manolo:
Regarding the light curve parameters (Fig.~\ref{paracut2}), an
increasing Doppler factor with distance ($d<0$) results into flatter
values of $\epsilon_{\rm flare\,amp.}$. This slope is slightly
steepening with increasing $b$ and $s$. The same trend with Doppler
factor evolution applies to $\epsilon_{\rm var.\,time}$, but this
parameter does not change significantly with $s$ or $b$, other than a
minor increase (implying a flattening of the slope). {Finally, the same
result is valid for $\epsilon_{\rm time\,lag}$, including some
irregular behaviours for $d=0$ and $d=-0.15$ at the edges of the studied parameter space, i.e., for $s\sim2.0$ and $b\sim2$ or for $s\sim3$ and $b\sim2$, see Fig. \ref{paraspace3}} The general trend is
that faster changes in the spectral evolution of the flare during the
Compton stage, as given by $d<0$ or $b\rightarrow 1$ (see
Fig.~\ref{paracut1}), result into smaller cross-band delays and {flare}
time-scales for the highest frequencies. For the low-frequency range
the time lags and differences in {flare} time-scales are also
smaller for $d<0$ (as it gives a flatter slope,
$\epsilon_\mathrm{time\,lag}$). {However}, in this case the slopes
are flatter for $b\rightarrow 2$, which gives a flatter
$\epsilon_\mathrm{time\,lag}$, too, because of steeper slopes during
the relevant stage {at} low frequencies {(i.e., the adiabatic
  stage, see Fig.~\ref{paracut1})}. Therefore, again, faster changes,
i.e., a steeper slope in the adiabatic stage, results into flatter
light curve parameters because of a faster passage of the peak
emission through the observing {frequencies}. On the contrary,
slower changes in the flare evolution along the jet provide more
differences among frequencies and thus, steeper light curve
parameters.

\subsection{General trends}
Our modelling of flares with the shock-in-jet model shows that the
evolution of the Doppler factor along the jet, $d$, has the largest
influence on the slopes of the different energy loss mechanisms.
Based on the assumption that the slope of the Compton stage has to be
$\epsilon_{\rm C}\leq0$ (i.e., {increasing} turnover flux density
with decreasing turnover frequency) as obtained from the observed
spectral evolution of these flares, we derive an upper limit for
$d=0.45$. With this upper boundary for $d$ we obtain a large range of
slopes $(0<\epsilon_{\rm C}<-14)$ for the initial stage of the
shock-in-jet model (assuming a symmetric lower limit of
$d=-0.45$). Therefore, jets with increasing Doppler factor along
the jet ($\delta\propto R^{-d}\rightarrow d<0$) can explain observed
steep Compton stages \citep[see, e.g.,][]{2013A&A...552A..11R}. Such
steep slopes, if considered as a general trend, can correspond to
acceleration of the bulk flow. The alternative explanation implies
that the flares in different jets approach the line of sight as they
evolve. The first option is favoured by the hypothesis of homogeneity,
which excludes privileged directions. We also find that jets with
purely toroidal magnetic field ($b=1$) develop steeper Compton stages
in the $\nu_\mathrm{m}-S_\mathrm{m}$~plane than jets with a purely poloidal field
($b=2$) (see panel A of Fig.~\ref{paracut1}). The spectral index of
the electron distribution does not play a crucial role on the slope of
the Compton stage.

In some single-dish observations of blazar flares the synchrotron
stage is not clearly visible or {apparently} missing
\citep[e.g.,][]{2011MmSAI..82..104T, 1996ApJ...466..158S} and some
authors claim that this stage may not {even} exist
\citep{2000ApJ...533..787B}. However, we {find} that for $d\neq0$
the slope of the synchrotron stage, $\epsilon_{\rm S}$ can be similar
to either the slope of the Compton stage ($d$<0) or the slope of the
adiabatic stage ($d$>0) (see~Fig.~\ref{vmsmplane1}). Due to the {
  usually} limited frequency sampling of single-dish {monitoring}
observations it is well possible that {a difference in the slopes
  between these two energy loss stages is not detected and} the
synchrotron stage is {just} ``hidden'' in the Compton or adiabatic
stage. The slope of the synchrotron stage in the $\nu_\mathrm{m} - S_\mathrm{m}$~plane
does not change significantly with the magnetic field, but it grows
with the spectral index of the electron distribution, becoming flatter
for negative values of $\epsilon_{\rm S}$, and steeper for positive
values.

Exponent $d$ also has a strong influence on the variation of the
light curve parameters. The time lag between any observed frequency
and a given reference frequency (345\,GHz in this work) can be used to
distinguish decelerating jets ($d>0$) from accelerating jets ($d<0$),
assuming constant viewing angle, $\vartheta=\mathrm{const}$. {Only for
$d>0$ we obtain time lags between the reference frequency of 345\,GHz
and frequencies $\nu\leq86$\,GHz which are in reach of typical cadence of single-dish monitoring programs ($>10$ days)}. 
This behaviour can be explained by the
slope of the Compton stage and the shape of the observed spectrum: the
optically thin flux density is given by
\begin{equation}
 S_{\nu,\mathrm{thin}}\propto \nu^{-(s-1)/2}.
\end{equation}
If the slope of the Compton stage, $\epsilon_{\rm C}$, is smaller {
  (i.e. steeper)} than $-(s-1)/2$, the optically thin flux density
{ at} a given frequency $\nu_i$ is still increasing after
$\nu_\mathrm{m}=\nu_i$. This behaviour is usually obtained for $d<0$
(accelerating jets, assuming constant viewing angle). However, if
$\epsilon_{\rm C}\sim-(s-1)/2$ the flux density peak { at} a given
frequency is equal to the { turnover point} ($\nu_\mathrm{m},\,S_\mathrm{m}$).

%Inaddition, the shape of the optically thin synchrotron spectrum is not
%a simple power law, as it also includes a cooling break, where the
%spectrum steepens. 
%As mentioned in Sect.~\ref{tsm}, we have modelled
%this break by including a fixed break frequency. If we assume a break
%frequency which is close to the turnover frequency, $\xi<5$, it is
%still possible to obtain high frequency time lags. However, this would
%require too {\bf strong} magnetic fields which are typically not
%expected in AGN jets {MP: QUANTIFY (give an example value of
%  the required magnetic field intensity)}. {\bf For a break frequency
%  as} assumed in this paper, no time lags are observed for $d>0$
%(decelerating flow, assuming constant viewing angle).

The cross-band delay exponent, $\epsilon_\mathrm{delay}$ is a measure of
the opacity variations along the jet {
\cite[e.g.,][]{1998A&A...330...79L,2011MNRAS.415.1631K,2014MNRAS.441.1899F}. 
For a conical jet with $b$=1 the expected value is
$\epsilon_\mathrm{delay}=1$, assuming equipartition between the
magnetic energy density and the energy density of relativistic
particles \citep{1998A&A...330...79L}. We obtain
$\epsilon_\mathrm{delay}=1$ for jets with $d\geq0$, $b<1.4$ and
$2.0<s<3.0$ (see panel A of Fig.~\ref{paracut2}, and
Fig.~\ref{paraspace3}). Those shock-models have large time lags
between low frequencies ($\nu<5$\,GHz) and the reference frequency
(here 345\,GHz) and show flat Compton stages ($\epsilon_{\rm
  C}>-2.0$). If we assume a fixed viewing angle those jets could be
classified as constant velocity or decelerating flows.

Both the standard shock-in-jet model for $d$>0 and the modified
Compton stage of \cite{2000ApJ...533..787B} give flat Compton
stages. This could be used to additionally distinguish between $d>0$
and $d<0$. However, second order Compton scattering could produce
non-zero time delays between the highest frequencies, too.

\subsection{A guide to identify shock parameters} 

{The work presented here can serve as (i) a guide to identify the jet
flow properties from the observed spectral evolution of flares in
blazar jets and (ii)} as a test for the validity of the shock-in-jet model
{(including or excluding second-order Compton scattering)}. 
Making use of {our} results and the previous discussion, we
provide here a short recipe for the extraction of shock parameters
from multi-frequency, single-dish light curves. {The frequency dependent light curve parameters, flare amplitude, flare time scale and the cross band delays can be obtained from well sampled light curves at any given frequency. If the flare of a blazar is observed by more than 2 frequencies, the exponents of the light curve parameters can be derived by fitting a power law or a broken power law (see, e.g., the flare amplitude for a broken power law) to the light curve parameters.  Given at least 5
frequencies and a relatively dense time sampling it is possible to extract the spectral evolution of a flare in the $\nu_\mathrm{m}-S_\mathrm{m}$ plane. Depending on the observing frequency range the entire spectral evolution can be obtained, i.e., covering the Compton, synchrotron and adiabatic stage, or in the case of a low frequency range only the synchrotron or the adiabatic stage. However, even in the case of only extracting one stage, based on its slope the evolution of the physical condition in the jet can be obtained.}
The frequency dependent light curve parameters (Sect.~\ref{slca}) can be
estimated by different methods, for instance, using
$\tau=t_{\mathrm{max,\nu_{i}}}-t_{\mathrm{max,ref}}$ (see Sect.~\ref{slca}) 
or a discrete cross-correlation function analysis 
\citep[e.g.][]{1988ApJ...333..646E,Larsson:2012vu} to obtain flare 
time lags.
Flare amplitudes can be obtained using e.g. light curve standard
deviations (see Sect.~\ref{slca}) or directly the minimum/maximum flux
density of the flare ($\Delta S=S_{\mathrm{max}}-S_{\mathrm{min}}$), {or using the intrinsic modulation index following \citet{2011ApJS..194...29R}}
The flare rise/decay times $\Delta t_{\mathrm{r,d}}$ can be
extracted using the times between the start, the maximum and the end
of the flare (see Sect.~\ref{slca}). Alternatively, each individual
(``isolated'') and well sampled flare in multi-frequency single dish
light curves could be approximated by}
\begin{eqnarray}
S_\nu&=&S_{\nu,0}+S_{\nu,\mathrm{max}}\exp{\left(t-t_0\right)/t_r}\qquad \mathrm{for\,} t<t_0\label{exp1}\\
S_\nu&=&S_{\nu,0}+S_{\nu,\mathrm{max}}\exp{-\left(t-t_0\right)/t_d}\qquad \mathrm{for\,}t>t_0,\label{exp2}
\end{eqnarray}
where $S_{\nu,0}$ is the quiescent flux for a given frequency,
$S_{\nu,\mathrm{max}}$ is the maximum flux density for a given
frequency, $t_0$ is the time at which the maximum flux density is
obtained, and $t_r$ and $t_d$ characterise the rise and decay times of
the flare, respectively \citep[see,
e.g.][]{2008A&A...485...51H,2013A&A...552A..11R}. {After modelling
  the flare at each frequency by Eq.~\ref{exp1} and \ref{exp2}}, the
frequency dependent parameters are computed as described in
Sect.~\ref{slca}.

By fitting a power law to the obtained frequency dependent light curve
parameters, the exponents $\epsilon_\mathrm{time\,lag}$,
$\epsilon_\mathrm{flare\,amp}$ and
$\epsilon_\mathrm{flare\,time\,scale}$ are obtained. The shock
parameters $d$, $s$ and $b$ can be estimated by comparing the
exponents $\epsilon_\mathrm{time\,lag}$, $\epsilon_\mathrm{flare\,amp}$
and $\epsilon_\mathrm{flare\,time\,scale}$ with {those} provided in
Figs.~\ref{paraspace3} or \ref{paraspace4}.

The relevant parameters can also be obtained from {studying} the
spectral evolution of the flare in the $\nu_\mathrm{m}-S_\mathrm{m}$--plane. This
additional analysis requires (quasi-) simultaneous {broad band
  spectra, constructed from (quasi-) simultaneous multi-frequency flux
  densities and/or an interpolation between observed data points}. In
this work we assumed that {the} total flux density is the
superposition {of} a quiescent and a flaring spectrum. Therefore
it is necessary to re-construct the quiescent spectrum from the
available or archival data. This quiescent spectrum could be modelled
by a simple power law $S_{\nu,q}\propto \nu^{\alpha_q}$ or by the full
expression for a {SSA} spectrum (see Eq.~\ref{snu}), depending on
the frequency sampling and the quality of the data. The evolution of
the turnover frequency, $\nu_\mathrm{m}$, the turnover flux density, $S_\mathrm{m}$, and
the optically thin spectral index, $\alpha_0$, are derived by fitting
Eq.~\ref{snu} to the {quiescent spectrum corrected flare spectra}.

The different energy loss stages can be identified in the
$\nu_\mathrm{m}-S_\mathrm{m}$--plane by their slopes, as described in
Sect.~\ref{tsm}. After identifying the energy loss stages and
extracting the slopes by means of a power law fit $S_{m,i}\propto
\nu_{m,i}^{\epsilon_i}$ \citep[see,][for details on the spectral
analysis of single dish light curves]{2011A&A...531A..95F}, the shock
parameters can be derived by comparing the obtained slopes for the
different stages with the ones provided in Figs.~\ref{paraspace1} and
\ref{paraspace2}.

\section{Conclusions and Outlook}
In this paper we used the shock-in-jet model and computed single-dish
light curves for different jet configurations. From the synthetic
light curves we extracted the slopes of the different energy loss
stages, the frequency-dependent light curve parameters which can be
used to distinguish different jet models and constrain the jet
parameters, e.g., $d$, $b$, and $s$.

Our results can be used to constrain the physical properties of jets
and their evolution within the collimation and acceleration regions,
by comparison with detailed {multi-frequency monitoring}
observations of blazar radio flares. As an example, a first comparison
between our analysis and observational results \citep[see,
e.g.,][]{2013A&A...552A..11R} could point towards bulk flow
acceleration during the initial Compton stage.
{Taking advantage of the broad-band (2.6 to 345\,GHz),
F-GAMMA monitoring program, we plan to apply this model to 
a large number of blazar flares to constrain} the jet and/or flare
properties. The parameters obtained can be used as initial
conditions in more advanced shock models, including the relativistic {magneto-hydrodyamic}
nature of jets to further investigate the physics of blazars. An
additional study for a parabolic jet geometry should be performed to
consider possible differences with conical jets that can be relevant
for the jet collimation region.

\begin{acknowledgements}
  We thank E. Ros for careful reading and useful comments on the
  manuscript. MP acknowledges financial support by the Spanish
  ``Ministerio de Ciencia e Innovaci\'on'' (MICINN) grants
  AYA2010-21322-C03-01, AYA2010-21097-C03-01. MP is a member of the work team of projects AYA2013-40979-P and AYA2013-48226-C3-2-P.
\end{acknowledgements}

\bibliographystyle{aa} 
\bibliography{biblio}

\section*{Appendix}
Here we present the results {for} the frequency dependent light curve
parameters {including} the entire range of shock parameters (see
Table~\ref{para}). The plots in this Appendix provide a global view of
the changes in the slopes of the different stages in the
$\nu_\mathrm{m}-S_\mathrm{m}$~plane and the light curve parameters.

\subsection*{Slopes of the energy loss stages}

Figures~\ref{paraspace1} and \ref{paraspace2} show the maps of the
values obtained for the slopes {of} the different energy loss
stages as a function of $d$, $b$, and $s$, indicating some colour
levels (black, solid lines) in the plot to help identifying the
values, and providing also a dashed line that separates the case of
evolution towards a magnetically dominated from evolution towards
particle dominated flows. This line is derived as follows: Taking into
account that the magnetic energy density is $u_B\propto B^2$ and using
$B\propto R^{-b}$, we obtain $u_B\propto R^{-2b}$. For particles,
$u_e\propto \int{n(\gamma)\gamma d\gamma}$ using $n(\gamma)\propto
K\gamma^-{s}$. Neglecting the evolution of $\gamma$ with distance, we
can assume $u_e\propto K$, and using $K\propto R^{-k}$ and
$k=2(s+2)/3$ if the jet expands adiabatically, we have $u_B/u_e\propto
R^{-2b+2(s+2)/3}$. Imposing independence with distance brings the
exponent to zero, which requires $b=(s+2)/3$. If $b <(s+2)/3$, the
ratio grows with distance, whereas for $b>(s+2)/3$ the ratio decreases
with distance. Each panel shows the variation of $\epsilon_i$ ($i=1$
Compton, $i=2$ synchrotron, and $i=3$ adiabatic) for $2<s<3$ and
$1<b<2$ and a fixed value of $d$. The value of $d$ is changing from
top to bottom from $d=-0.45$ to $d=0.45$ (see also the figure
captions). The left column in both plots shows the maps of values of
$\epsilon_{\rm C}$ as a function of $b$ and $d$. The vertical levels
indicate that this slope is fairly independent of $s$ for any values
of $b$ and $d$, and that it mainly changes with these two
parameters. In the case of $\epsilon_{\rm S}$, the slope of the
synchrotron stage, the situation is different, and $s$ and $d$ appear
to be the most relevant parameters to determine it, although there is
also a smooth gradient of this slope in the direction of $b$ for the
extreme values of $d$. Finally, the third column shows the maps of
$\epsilon_{\rm A}$, which is most sensitive to $d$ and $b$, and only
shows a smooth variation with $s$.

\begin{figure*}
\centering
\includegraphics[width=17cm]{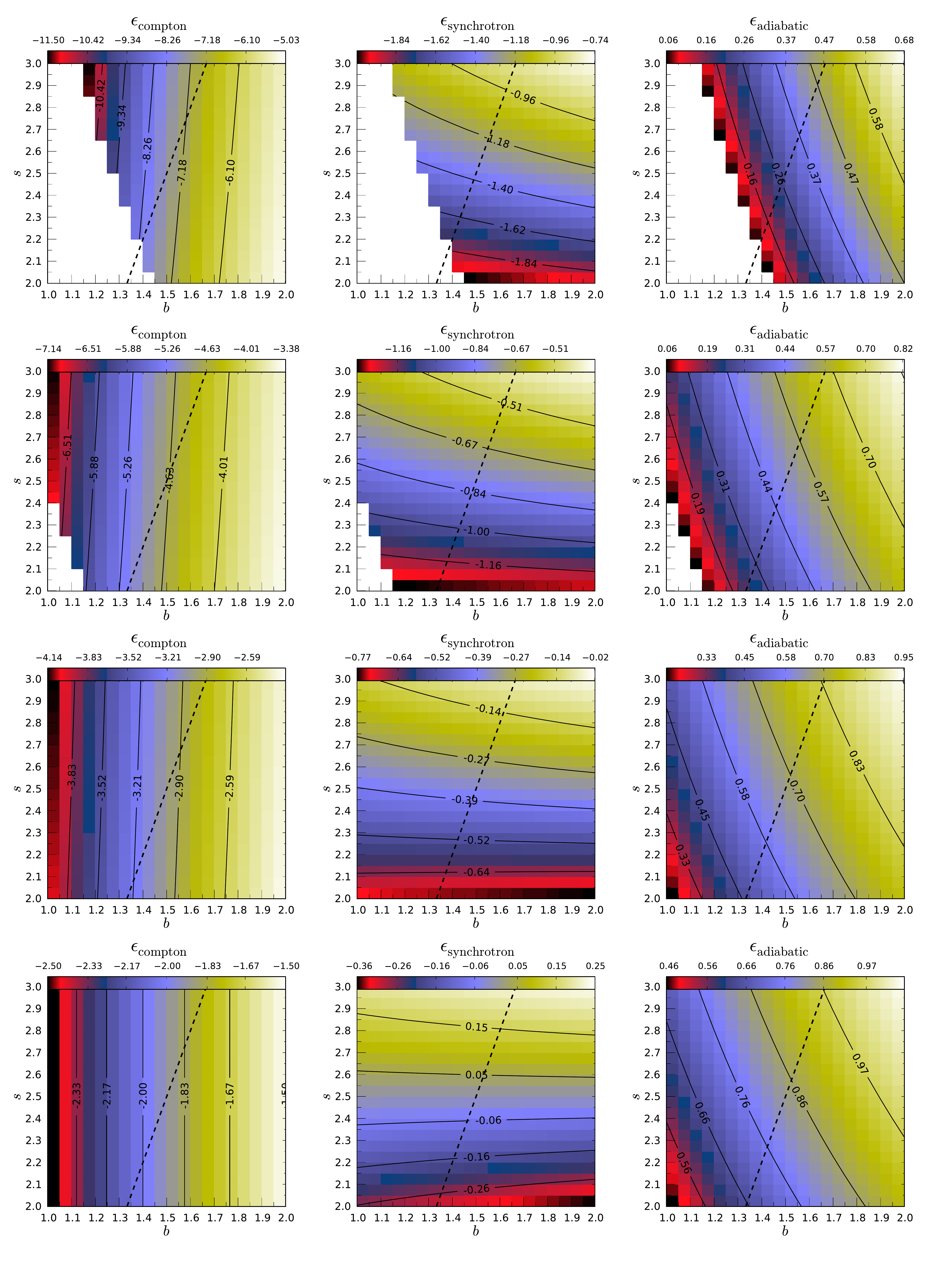}
 \caption{Parameter space plots for the variation of the slopes,
       $\epsilon_i$ as function of $b$ and $s$ while {keeping the $d$ parameter fixed.}
       The columns show from left to right, the slope
       of the Compton stage, $\epsilon_{\rm C}$, the slope of the
       synchrotron stage, $\epsilon_{\rm S}$, and the slope of the
       adiabatic stage, $\epsilon_{\rm A}$. The exponent for the
       evolution of the Doppler factor, $d$, is from top to bottom
       $d=-0.45$, $d=-0.30$, $d=-0.15$, and $d=0$. The black dashed line
       corresponds to a constant $u_B/u_e$ ratio with distance
       ($b_\mathrm{eq}=(s+2)/3)$), i.e., to the left of this line the
       jet flow tends to be magnetically dominated with distance and
       to the right the jet tends to be particle energy dominated with
       distance.}
\label{paraspace1} 
\end{figure*}

\begin{figure*}
\centering
\includegraphics[width=17cm]{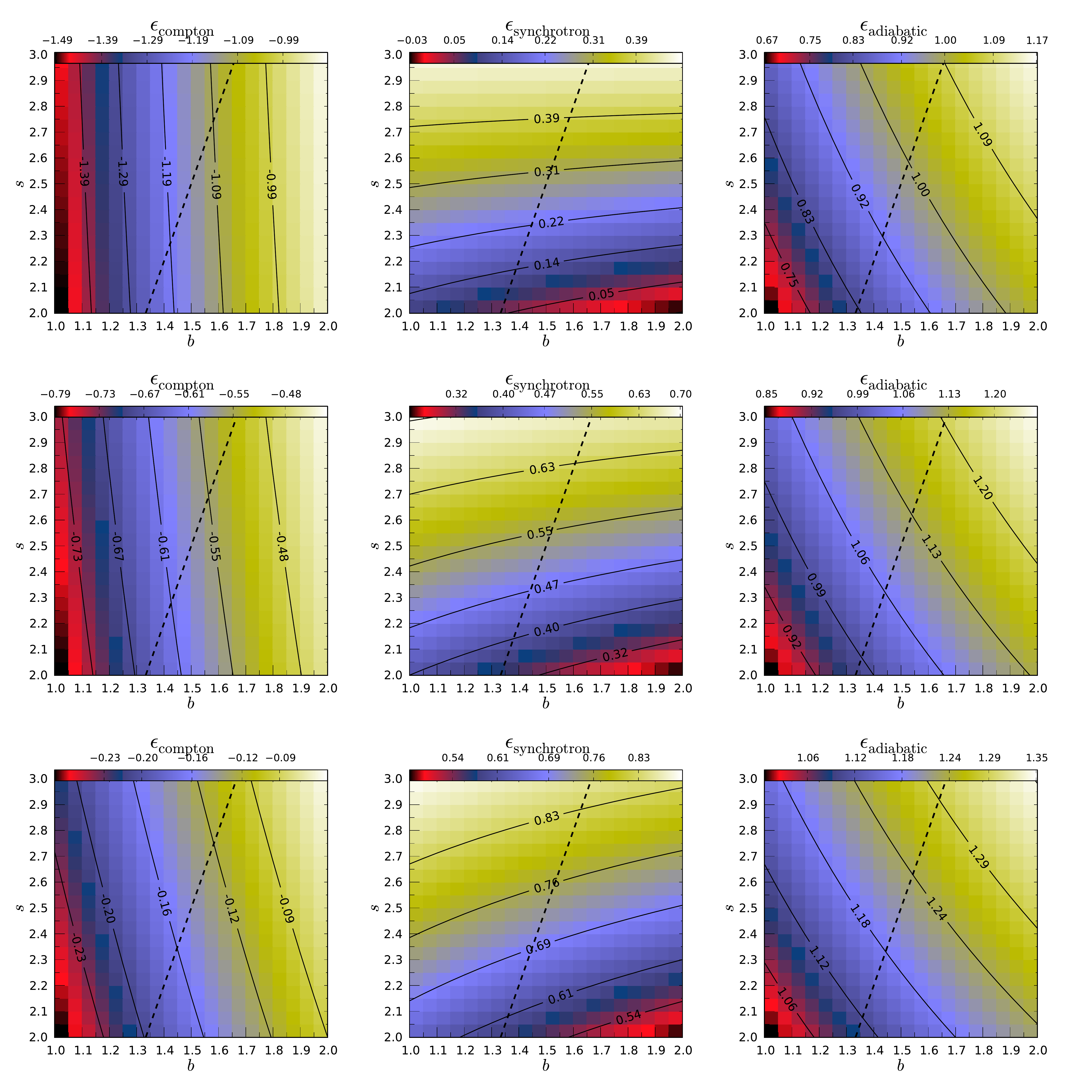}
 \caption{Same as Fig .\ref{paraspace1} for $d=0.15$, $d=0.30$, and $d=0.45$.}
\label{paraspace2} 
\end{figure*}

\subsection*{Slopes of the frequency dependent light curve parameters}
Figures~\ref{paraspace3} -- \ref{paraspace6} show maps of the
exponents of the frequency dependent light curve parameters as a
function of $s$ and $b$ for different values of $d$. 
{As mentioned earlier, the exponents for the flare amplitude and the flare time scale
can be obtained either from the rising edge or the decaying edge of the light curve. In Fig \ref{paraspace3} and \ref{paraspace4}
we present these parameters obtained from the rising edge and in Fig \ref{paraspace5} and \ref{paraspace6} for the decaying edge of the light curves.}
In Fig \ref{paraspace3} and \ref{paraspace4}, each panel shows,
from left to right, the variation in the flare amplitude exponent,
$\epsilon_\mathrm{flare\,amp.}$, the flare time scale
exponent $\epsilon_\mathrm{flare\,time\,scale}$ and the cross-band delay exponent,
$\epsilon_\mathrm{delay}$, for $2<s<3$ and $1<b<2$
and a fixed value of $d$. The value of $d$ is changing from top to
bottom from $d=0.45$ to $d=-0.45$ (see also the figure captions). The
amplitude of the flare undergoes a stronger variation with frequency
for decreasing Doppler factors with distance (Fig.~\ref{paraspace3})
than for the increasing (Fig.~\ref{paraspace4}), as indicated by the
colour-scales. In all cases, the slope grows with increasing $s$ and
$b$. The time lapse between the onset of the flare and the peak at
each frequency is more sensitive to changes in frequency for
decreasing Doppler factors with distance. This time lapse is more
sensitive to $b$ than to $s$, and the difference among frequencies
becomes larger (smaller $\epsilon_{\rm flare\,time\,scale}$) for values
of $b$ closer to 1. Finally, the time lag between the peaks at
different frequencies and a reference one has a similar behavior with
respect to the relevant parameters to the the time lapse between
onsets and peaks. The main difference is that there is not a large
difference in the slopes between positive and negative values of $d$
and that there are clear discontinuities in the values of the
$\epsilon_{\rm time\,lag}$ for increasing Doppler factors, at certain
values of $s$.

{In Fig. \ref{paraspace5} and \ref{paraspace6} we show the variation of the exponent for 
the flare amplitude and the flare time scale obtained from the decaying edge of the light curve.
The exponent for the flare amplitude $\epsilon_\mathrm{flare\,amp.\,decay}$ decreases with $d$.
For $d<0$ the absolute value of the exponent increases with $s$ and $b$. However, for $d>0$ the 
distribution of $\epsilon_\mathrm{flare\,amp.\,decay}$  changes: The exponents still increase with $s$ but larger values
are obtained towards $b=1$. The exponent for the flare time scale derived from the decaying edge of the light curve, $\epsilon_\mathrm{flare\,time\,decay}$ is small, typically $<0.05$. The value and its distribution depend strongly on $d$. For $d<0$ the distribution is smooth and the values decrease with $s$ and $b$. Nearly no variation in $\epsilon_\mathrm{flare\,time.\,decay}$ is obtained for $d>0$ (see second column in Fig. \ref{paraspace5} and \ref{paraspace6}).}

Figures~\ref{paraspace7} and \ref{paraspace8} show the expected time
lags (in years) between the peaks at 5, 15, and 140~GHz and our
reference frequency, 345~GHz (left, central and right columns,
respectively), for different values of $d$ (different rows), as a
function of $s$ and $b$. The Cross-band delays become shorter for increasing
Doppler factor with distance as indicated by the colour scales at the
top of the panels. The time lags between the reference frequency and low
frequencies are typically more sensitive to $b$ increasing as this
parameter tends to 1, whereas the time lags between 140 and 345~GHz show
significant values only for decreasing Doppler factor with distance
and higher sensitivity to the spectral slope $s$.

\begin{figure*}
\centering
\includegraphics[width=17cm]{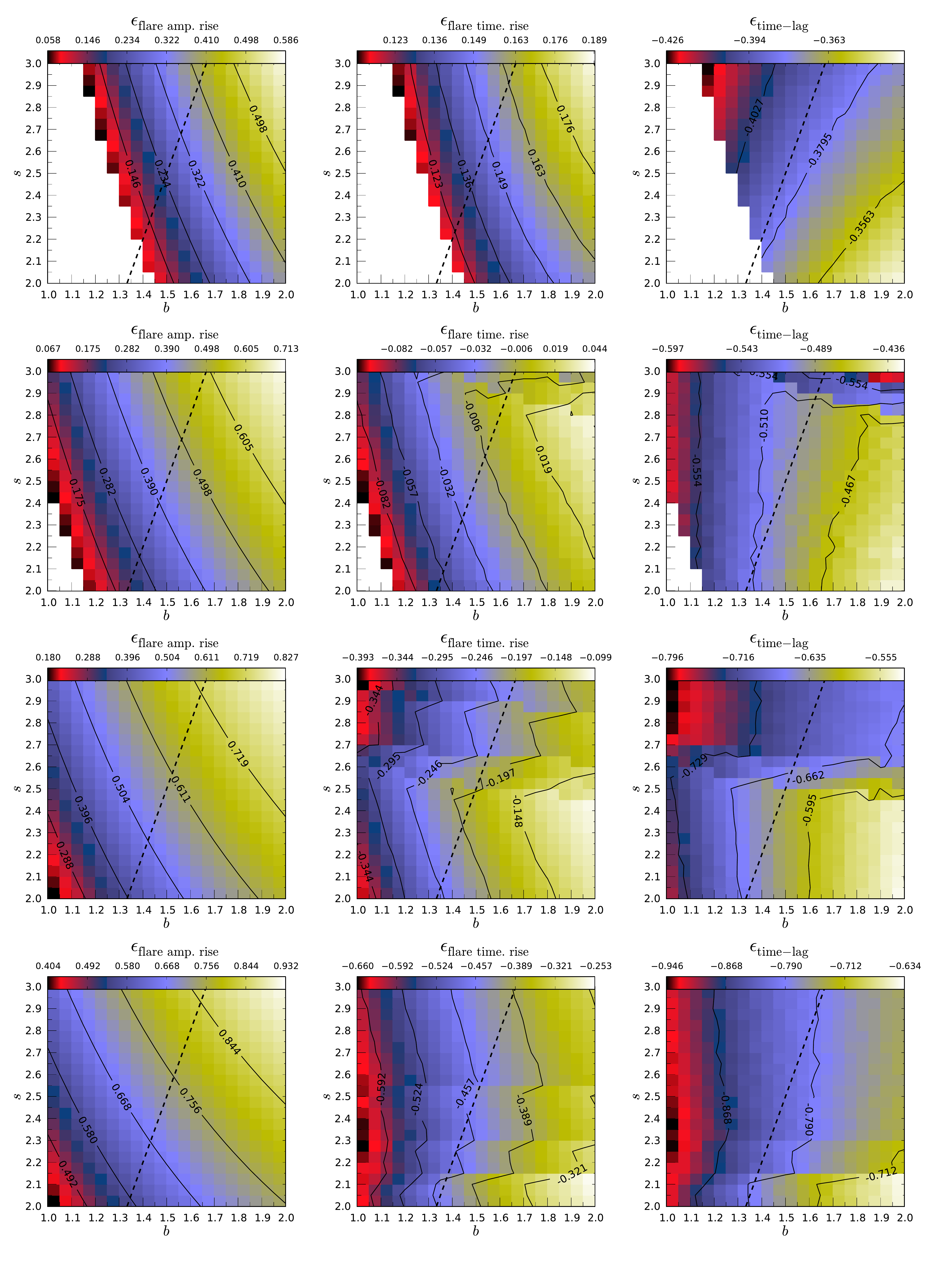}
 \caption{Parameter space plots for the variation of
       frequency-dependent single-dish light curve parameters {obtained from the rising edge of the light curves} as
       function of $b$ and $s$ while {keeping the $d$ parameter fixed.}
       The columns show from left to right the exponent for the
       variability amplitdue, $\epsilon_\mathrm{flare\,amp.}$, the
       exponent for the variability time scale, $\epsilon_\mathrm{flare\,time\,scale}$,
        and the exponent for the time lag, $\epsilon_\mathrm{delay}$.
       The exponent for the
       evolution of the Doppler factor, $d$, is from top to bottom
       $d=-0.45$, $d=-0.30$, $d=-0.15$, and $d=0$. The black dashed line
       corresponds to a constant $u_B/u_e$ ratio with distance
       ($b_\mathrm{eq}=(s+2)/3)$), i.e., to the left of this line the
       jet flow tends to be magnetically dominated with distance and
       to the right the jet tends to be particle energy dominated with
       distance.}
\label{paraspace3} 
\end{figure*}
\newpage

\begin{figure*}
\centering
\includegraphics[width=17cm]{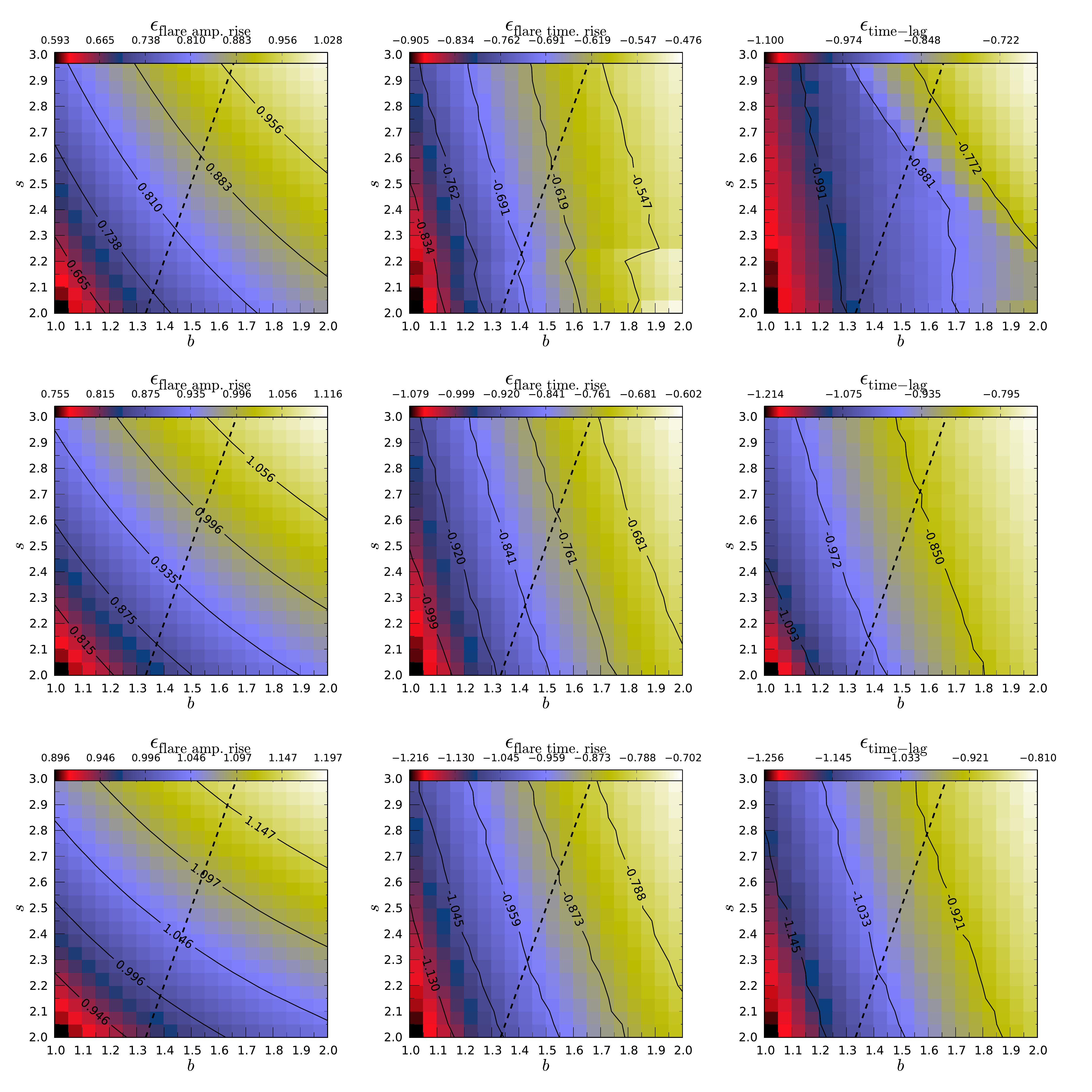}
   \caption{Same as Fig.~\ref{paraspace3} for $d=0.15$, $d=0.30$, and $d=0.45$.}
\label{paraspace4} 
\end{figure*}
\newpage

\begin{figure*}
\centering
\includegraphics[width=17cm]{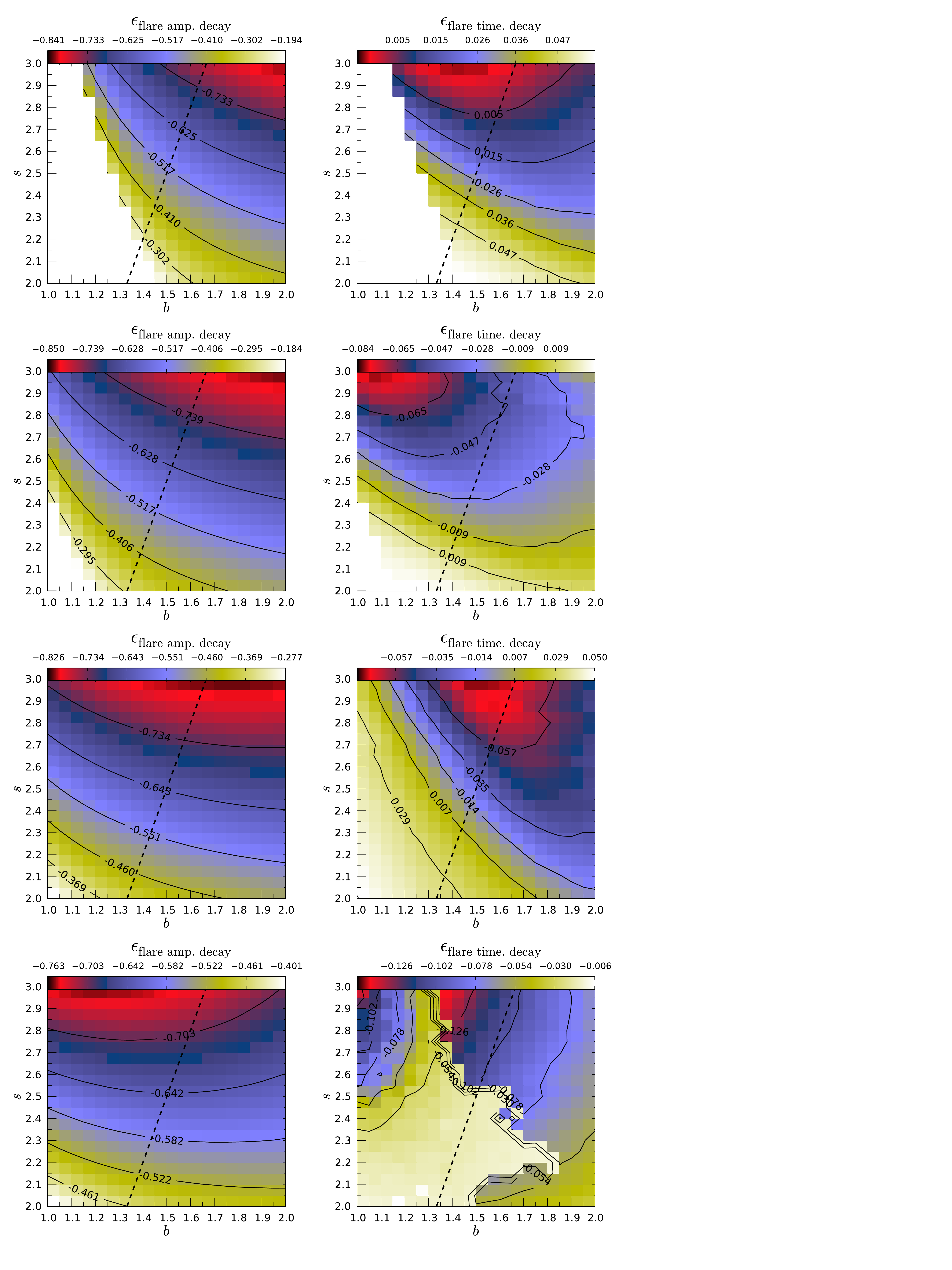}
 \caption{Parameter space plots for the variation of
       frequency-dependent single-dish light curve parameters {obtained from the decaying edge of the light curves} as
       function of $b$ and $s$ while {keeping the $d$ parameter fixed.}
       The columns show from left to right, the exponent for the
       variability amplitdue, $\epsilon_\mathrm{var.\,amp.}$, the
       exponent for the variability time scale,
       $\epsilon_\mathrm{var.\,time\,scale}$, and the exponent for the
       cross-band delay, $\epsilon_\mathrm{delay}$. The exponent for the
       evolution of the Doppler factor, $d$, is from top to bottom
       $d=-0.45$, $d=-0.30$, $d=-0.15$, and $d=0$. The black dashed line
       corresponds to a constant $u_B/u_e$ ratio with distance
       ($b_\mathrm{eq}=(s+2)/3)$), i.e., to the left of this line the
       jet flow tends to be magnetically dominated with distance and
       to the right the jet tends to be particle energy dominated with
       distance.}
\label{paraspace5} 
\end{figure*}
\newpage

\begin{figure*}
\centering
\includegraphics[width=17cm]{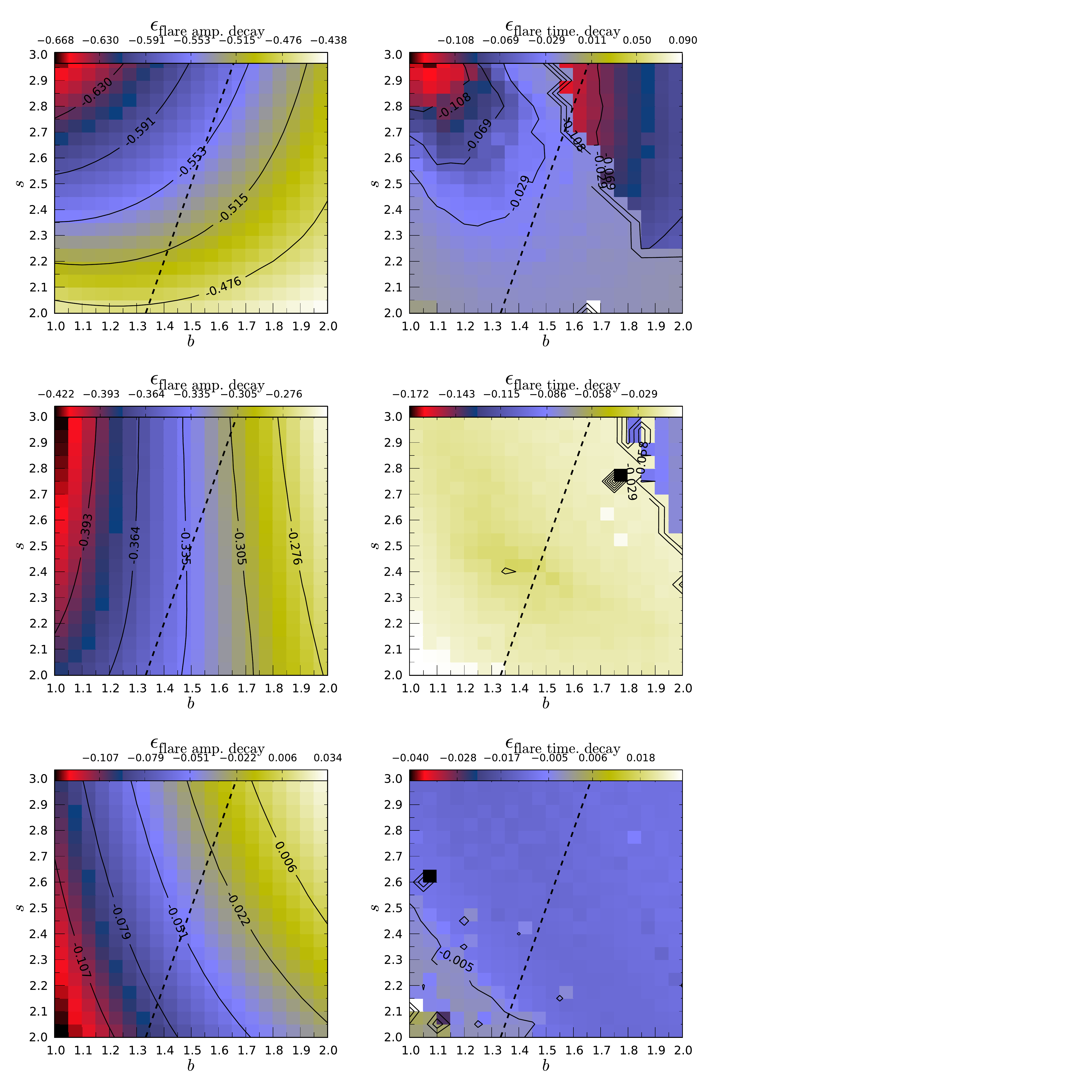}
   \caption{Same as Fig.~\ref{paraspace3} for $d=0.15$, $d=0.30$, and $d=0.45$.}
\label{paraspace6} 
\end{figure*}
\newpage

\begin{figure*}
\centering
\includegraphics[width=17cm]{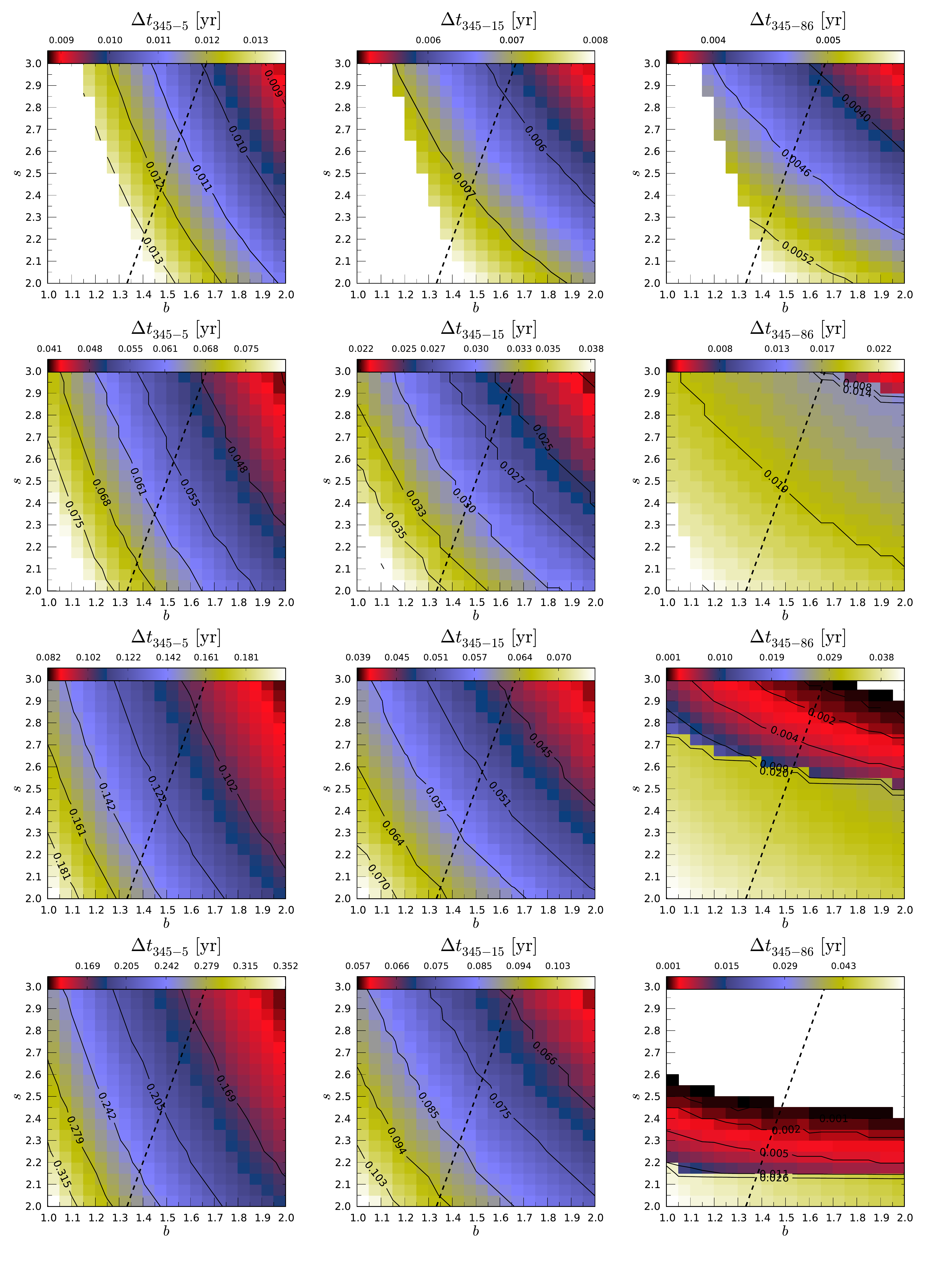}
     \caption{Parameter space plots for the for the time lag between
       three selected frequencies as function of $b$ and $s$ while
       {keeping the $d$ parameter fixed.} The columns show from left to
       right, the (345-5)GHz time lag, the (345-15)GHz time lag, and
       the (345-86)GHz time lag. The exponent for the evolution of
       the Doppler factor, $d$, is from top to bottom $d=-0.45$,
       $d=-0.30$, $d=-0.15$, and $d=0$. The black dashed line
       corresponds to a constant $u_B/u_e$ ratio with distance
       ($b_\mathrm{eq}=(s+2)/3)$), i.e., to the left of this line the
       jet flow tends to be magnetically dominated with distance and
       to the right the jet tends to be particle energy dominated with
       distance.}
\label{paraspace7} 
\end{figure*}
\newpage

\begin{figure*}
\centering
\includegraphics[width=17cm]{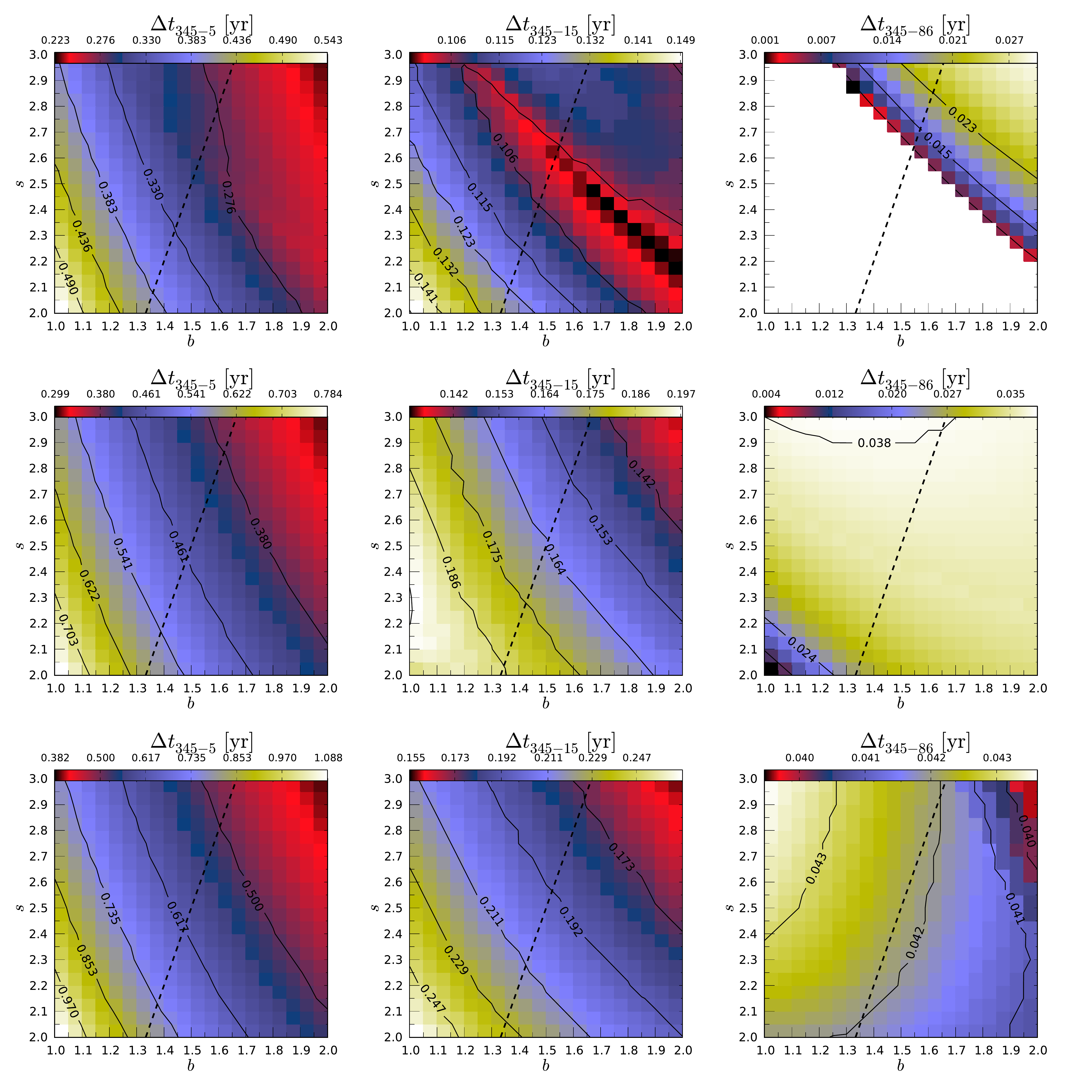}
\caption{Same as Fig.\ref{paraspace7} for $d=0.15$, $d=0.30$, and $d=0.45$.}
\label{paraspace8} 
\end{figure*}
\newpage

\begin{figure*}
\centering
\includegraphics[width=16cm]{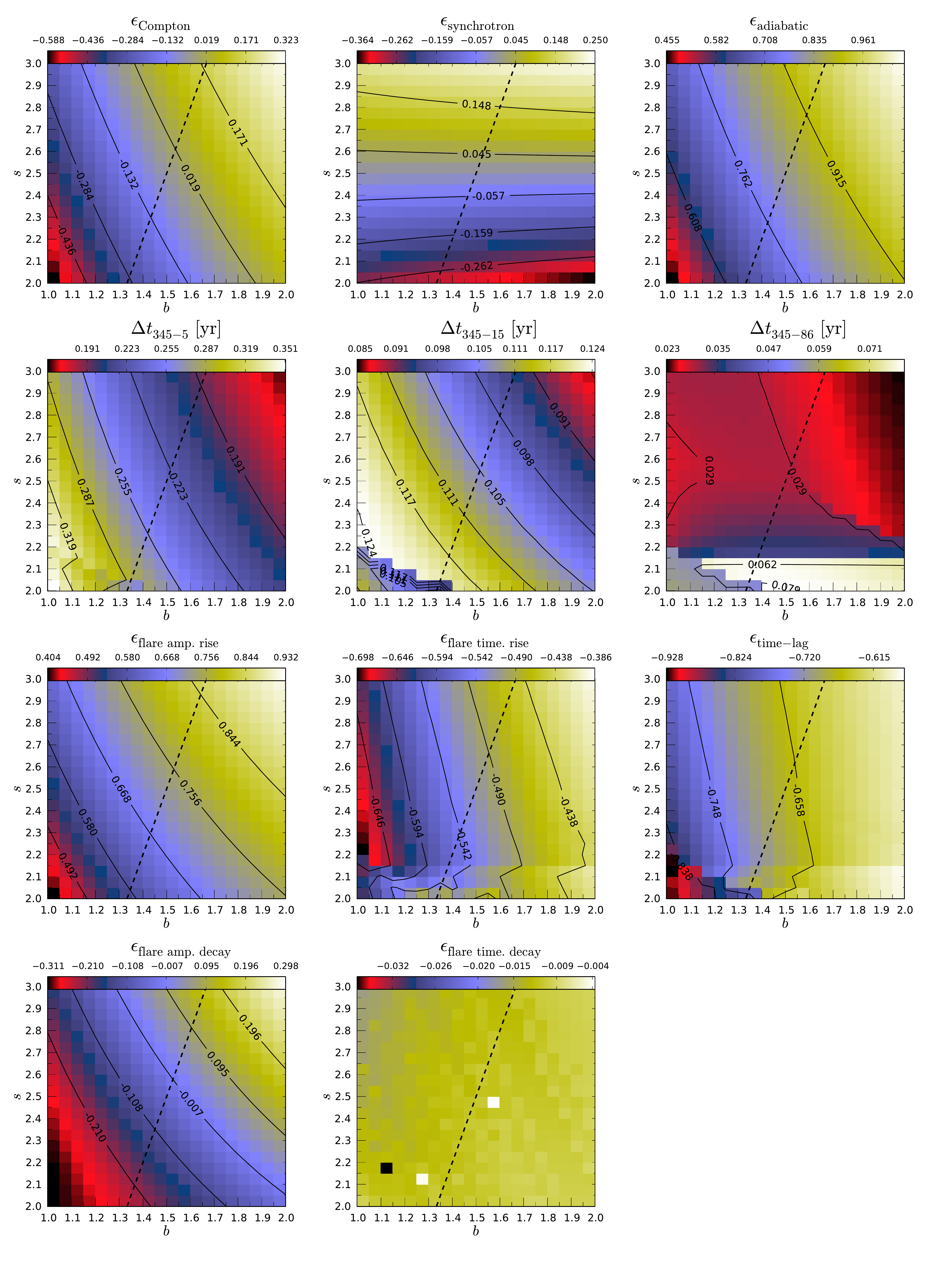}
     \caption{Modified Compton stage model using $d=0$. The top row
       shows the slopes for the different energy loss stages from left
       to right: Compton, synchrotron, and adiabatic stage. The delay between
       selected frequencies with respect to 345\,GHz in years is
       plotted in the second row from left to right: delay to 5GHz,
       delay to 15\,GHz and delay to 86\,GHz. The third row
       presents the frequency-dependent light curve parameters obtained
       from the rising edge of the light curve
       from left to right: flare amplitude, flare time scale 
       and cross frequency delay. The bottom row shows the exponent for the flare amplitude and the 
       flare time scale as derived from the decaying edge of the light curve.   
       The black dashed line
       corresponds to a constant $u_B/u_e$ ratio with distance
       ($b_\mathrm{eq}=(s+2)/3)$), i.e., to the left of this line the
       jet flow tends to be magnetically dominated with distance and
       to the right the jet tends to be particle energy dominated with
       distance.}
\label{modcompton} 
\end{figure*}

\end{document}